\documentclass[12pt,epsf]{article}

\begin{document}

\baselineskip=18.8pt plus 0.2pt minus 0.1pt

%%%%%%%%%%% Private Macros %%%%%%%%%%%%%
\makeatletter

%\@addtoreset{equation}{section}
%\renewcommand{\theequation}{\thesection.\arabic{equation}}
\renewcommand{\thefootnote}{\fnsymbol{footnote}}
\newcommand{\beq}{\begin{equation}}
\newcommand{\eeq}{\end{equation}}
\newcommand{\bea}{\begin{eqnarray}}
\newcommand{\eea}{\end{eqnarray}}
\newcommand{\nn}{\nonumber \\}
\newcommand{\hs}[1]{\hspace{#1}}
\newcommand{\vs}[1]{\vspace{#1}}
\newcommand{\Half}{\frac{1}{2}}
\newcommand{\p}{\partial}
\newcommand{\ol}{\overline}
\newcommand{\wt}[1]{\widetilde{#1}}
\newcommand{\wh}[1]{\widehat{#1}}
\newcommand{\ap}{\alpha'}
\newcommand{\bra}[1]{\left\langle  #1 \right\vert }
\newcommand{\ket}[1]{\left\vert #1 \right\rangle }
\newcommand{\vev}[1]{\left\langle  #1 \right\rangle }
\newcommand{\vac}{\ket{0}}

\newcommand{\ul}[1]{\underline{#1}}

\makeatother
%%%%%%%%% End of private macros %%%%%%%%%%%

\begin{titlepage}
\title{
\hfill\parbox{4cm}
{\normalsize MIT-CTP-3724\\{\tt hep-th/0602251}}\\
\vspace{1cm}
Field Redefinitions, T-duality and Solutions in Closed String Field Theories
}
\author{Yoji Michishita
\thanks{
{\tt michishi@lns.mit.edu}
}
\\[7pt]
{\it Center for Theoretical Physics}\\
{\it Massachusetts Institute of Technology}\\
{\it Cambridge MA 02142 USA}
}

\date{\normalsize February, 2006}
\maketitle
\thispagestyle{empty}

\begin{abstract}
\normalsize
We investigate classical solutions in closed bosonic string field theory and 
heterotic string field theory that are obtained order by order
starting from solutions of the linearized equations of motion,
and we discuss the ``field redefinitions'' which relate massless fields
on the string field theory side and the low energy effective theory side.
Massless components of the string field theory solutions are not
corrected and from them we can infer corresponding solutions in the
effective theory: the chiral null model and the pp-wave solution
with B-field, which have been known to be $\ap$-exact. These two sets of
solutions on the two sides look slightly different because of the field
redefinitions. It turns out that T-duality is a useful tool to determine them:
We show that some part of the field redefinitions can be determined
by using the correspondence between T-duality rules on the two sides,
irrespective of the detail of the interaction terms and the integrating-out
procedure.
Applying the field redefinitions, we see that the solutions on the
effective theory side are reproduced from the string field theory solutions.
\end{abstract}

%PACS codes: 11.25.-w

%Keywords: string field theory

\end{titlepage}

%%%%%%%%%%%%%%%%%%%%%%%%%%%%%%%%%%%%%%%%%%%%%%%%%%%%%%%%%%%%%%%%%%%
%%%%%%%%%%%%%%%%%%%%%%%%%%%%%%%%%%%%%%%%%%%%%%%%%%%%%%%%%%%%%%%%%%%
\section{Introduction}

In closed string field theories, which are nonpolynomial analogs of Witten's
cubic open string field theory\cite{w}, there are few known examples of
classical
solutions. On the other hand, in the low energy effective field theories
we know many solutions. It is important to fill this gap because closed
string field theories may give nonperturbative formulations of string theory.

We have an additional complication when we compare solutions in closed string
field theories with those in the effective theories:
Massless components of string fields are related to massless fields in
the low energy effective theory. However, their relation is not direct
except at the leading order. They are related by some complicated 
``field redefinitions''. Since massless fields in the effective theory or
gauge invariant quantities made of them have direct physical and
geometrical meaning, it is very important to investigate these field
redefinitions.

In this paper we discuss the field redefinitions, and applying it we
investigate some classical solutions in the bosonic closed
string field theory constructed in \cite{z1,csft}, and in the heterotic
string field theory constructed in \cite{oz,boz}.
We construct the solutions in a way similar to \cite{mchsht}:
They are constructed order by order
starting from solutions of the linearized equations of motion. We have to
introduce appropriate source terms in the equations to obtain some solutions.
It can be shown that the massless components of these solutions do not
receive higher order corrections. Roughly speaking, this is because
our solutions contain only $\alpha_{-m}^-$ and no $\alpha_{-m}^+$, and
therefore higher order corrections have more and more $\alpha_{-m}^-$.
On the other hand, on the effective field theory side similar solutions
have been known: the chiral null model and the pp-wave solution with B-field.
They are $\ap$-exact solutions\cite{ht,cmp,tstl,dhh}, and can also be
regarded as
nonlinear extensions of solutions of the linearized equations of motion.
Since linearized equations of motion of both sides are equivalent, and
the nonlinear extensions of the linearized solutions look very similar,
it is natural to identify them as the same solutions. However, the solutions
look slightly different because of the effect of the field redefinitions.

Therefore we need explicit expressions of the field redefinitions for
comparison between those solutions.
We show that some part of them can be determined by using
T-duality transformation. Since we know how T-duality transforms fields on
both the string field theory side and the effective theory side at least in
the lowest order in $\ap$, the field redefinitions are restricted by
the correspondence between two T-duality rules. Although this does not
fully determine them, we stress that this method does not depend
on the detail of the interaction terms and the procedure of
integrating out massive fields, and therefore the result is universal.
Using the field redefinitions and assuming that
higher derivative terms in them cancel, we see that our string field theory
solutions really correspond to those in the effective theory.

This paper is organized as follows:
In section 2, we investigate solutions in open string field theories
as a preliminary to the closed string case.
These solutions give simplified version of those introduced in later sections,
and are interesting in its own right.
In section 3, we review basic facts about the closed bosonic string field
theory and the heterotic string field theory, fix notation, and give
some general argument about the field redefinitions.
In section 4, we determine some terms in the field redefinitions
by using the correspondence between T-duality rules on the string field theory
side and the effective theory side.
In section 5, we give string field theory solutions which are constructed by
the same way as in section 2, and have properties similar to them.
We can find corresponding solution on the effective theory side which
is known to be $\ap$-exact: the chiral null model. We confirm the
correspondence by applying the field redefinitions.
In section 6, we give more solutions which again have properties similar to 
the one in section 2, find corresponding solution on the effective theory side
which is known to be $\ap$-exact under some condition: the pp-wave solution
with B-field, and confirm the correspondence.
Section 7 contains some discussion.

Let us give some remark about notation in this paper.
In the effective theory we always use string metric.
We put hats on quantities in the effective theory, and corresponding
quantities in the string field theory are denoted by the same symbols
without hats. Spacetime indices denoted by $*$ may be free or may be
contracted with others.
Unless otherwise noted, spacetime indices are raised and lowered by
$\eta_{\mu\nu}$.

%%%%%%%%%%%%%%%%%%%%%%%%%%%%%%%%%%%%%%%%%%%%%%%%%%%%%%%%%%%%%%%%%%%
%%%%%%%%%%%%%%%%%%%%%%%%%%%%%%%%%%%%%%%%%%%%%%%%%%%%%%%%%%%%%%%%%%%
\section{Solutions in Open String Field Theories}

In this section we give classical solutions in open string field theories on
one single D$p$-brane in the flat spacetime. The structure of these solutions
is similar to the closed string field configuration in section 5,
and will help readers understand more complicated closed string cases.

For definiteness we use Witten's bosonic cubic string field theory\cite{w}
and Berkovits' superstring field theory\cite{bkv}. However our discussion
does not depend on the detail of the interaction vertices as long as they
are defined by using correlators of conformal field theory.

We separate spacetime coordinates $x^\mu$ into
$x^\pm=\frac{1}{\sqrt{2}}(x^0\pm x^1)$ and $x^i$.
$x^\pm$ and some of $x^i$ are along the D-brane.
$\wh{A}_\mu$ denote gauge fields or scalar fields in the low energy effective
field theory on the D$p$-brane, depending on $\mu$.

We put the ansatz
$\wh{A}_+=0$, $\wh{A}_i=0$ and $\wh{A}_-=\wh{A}_-(x^-,x^i)$.
Then the linearized equation of motion for $\wh{A}_\mu$ reduces to
\beq
\p_i\p^i\wh{A}_-(x^-,x^i)=0.
\label{oefflineq}
\eeq
By solving this equation we obtain
\beq
\wh{A}_-(x^-,x^i)=\sum_I\frac{c_I(x^-)}{[\sum_i(x^i-x^i_I)^2]^{(p-3)/2}}
 +d_i(x^-)x^i+f(x^-),
\label{oeffsol}
\eeq
where $x^i_I$ are constants and $c_I(x^-)$, $d_i(x^-)$ and $f(x^-)$ are
arbitrary functions of $x^-$. We assumed $p\geq 4$. (Of course there are more
solutions, such as $d_{ij}(x^-)x^ix^j$ with $d_i^{\;\; i}=0$.)
$f(x^-)$ can be gauged away without changing the ansatz.
The first term in (\ref{oeffsol}) is analogous to the solution in \cite{cm},
and for this term we have to introduce delta function source terms in
(\ref{oefflineq}).
The second term is better known in the following gauge transformed form:
\bea
\wh{A}_- & \rightarrow & \wh{A}_-
 +\p_-\left[-x^i\int^{x^-}dx'^-d_i(x'^-)\right]=0, \\
\wh{A}_+ & \rightarrow & \wh{A}_+=0, \\
\wh{A}_i & \rightarrow & \wh{A}_i
 +\p_i\left[-x^i\int^{x^-}dx'^-d_i(x'^-)\right]=\int^{x^-}dx'^-d_i(x'^-).
\eea
In other words, $\wh{A}_\pm=0$ and $\wh{A}_i$ are arbitrary functions of $x^-$.
This configuration and its T-dualized ones have been investigated in
\cite{ck,bh,st,bch}, and have been shown to be $\ap$-exact solutions.

We can construct string field theory version of this solution order by order,
in the same way as that of \cite{mchsht}.
We can give an string field theory proof of the $\ap$-exactness.
First let us consider bosonic case.

Notice that the string field configuration
\beq
\Phi_0=\int\frac{d^{26}k}{(2\pi)^{26}}iA_\mu(k)c\p X^\mu e^{ik\cdot X}
\eeq
with $A_+=0$, $A_i=0$, $A_-=A_-(k_-,k_i)$ and $k_ik^iA_-=0$,
satisfies the linearized
string field equation $Q\Phi_0=0$, which is equivalent to the linearized
equation of motion of the effective theory.\footnote{
Throughout this paper we freely switch from coordinate expression to
momentum expression, and vice versa. They are related by Fourier
transformation. $A_-=A_-(k_-,k_i)$ really means that $A_-(k)$ is equal to
delta function $\delta(k_+)$ times a function depending on $k_-$ and $k_i$.
We hope this kind of notation causes no confusion.}

The full order solution is constructed by expanding the string field $\Phi$
in some parameter $g$:
\beq
\Phi=g\Phi_0+g^2\Phi_1+g^3\Phi_2+\dots.
\eeq
The fully nonlinear equation of motion
\beq
Q\Phi+\Phi^2=0 \label{obsfteom}
\eeq
is decomposed into contributions from each order in $g$:
\beq
Q\Phi_n+\sum_{m=0}^{n-1} \Phi_m\Phi_{n-m-1}=0.
\eeq
Then imposing the condition $b_0\Phi_n=0$, we can solve these equations:
\beq
\Phi_n=-\frac{b_0}{L_0}\sum_{m=0}^{n-1} \Phi_m\Phi_{n-m-1}.
\eeq
If we want to obtain the solution corresponding the first term of 
(\ref{oeffsol}), we have to introduce the source term in (\ref{obsfteom}).
For that case see \cite{mchsht}.

$\Phi_n$ has no $k_+$ dependence, and consists of states with
$n_--n_+\geq n+1$, where $n_\pm$ are numbers of $\alpha^\pm_{-m}$ in the Fock
space representation of $\Phi_n$. This can be proven by almost the same
argument as in \cite{mchsht}.
This simple fact leads us to many nice properties:
$\Phi$ has no tachyon component, and the massless component has no higher
order contribution. Therefore inverses of $L_0$ are well-defined.
Nonzero coefficient of each Fock space state receives
contribution from finite number of $\Phi_n$.

In general, $A_\mu$ is different from the gauge field of the effective
theory $\wh{A}_\mu$, because their gauge transformations are different.
We need ``field redefinition'' to relate them. A procedure to compute it
order by order has been explained in \cite{cst}.
(See also \cite{d}.)
$\wh{A}_\mu$ is expressed as a functional of $A_\mu$ as follows:
\beq
\wh{A}_\mu=A_\mu+\mbox{(terms which are quadratic or higher in $A_\mu$ and
 may have derivatives)}.
\label{aredef}
\eeq
Here tachyon component is regarded as a massive field and integrated out,
or is just put zero after we determine the form of the field redefinition
involving with the tachyon. Since for our solution the tachyon component
is exactly zero, either way eventually lead us to the same conclusion.

Fortunately for our solution, $\wh{A}_\mu$ is equal to $A_\mu$,
because the correction terms in (\ref{aredef}) contain either $\p_\mu A^\mu$ or
$A_\mu A^\mu$, which are zero for our solution. Combining this with the fact
that $A_\mu$ has no higher order correction from $\Phi_n$ with $n\geq 1$,
we can see that our $A_\mu=\wh{A}_\mu$ is an $\ap$-exact solution.

We can construct a similar solution in Berkovits' superstring field theory.
The lowest order solution $\Phi_0$ is given by
\beq
\Phi_0=\int\frac{d^{10}k}{(2\pi)^{10}}
 A_\mu(k)\xi c\psi^\mu e^{-\phi}e^{ik\cdot X},
\eeq
where $A_+=0$, $A_i=0$ and $A_-=A_-(k_-,k_i)$, and $A_-$ satisfies
$k_ik^iA_-=0$.
The equation of motion is
\bea
0 & = & \eta_0(e^{-\Phi}Qe^\Phi) \nn
 & = & \sum_{n=0}^\infty\frac{(-1)^n}{(n+1)!}
 \eta_0[\underbrace{\Phi,[\Phi,[\dots,[\Phi}_{n},Q\Phi]]\dots ].
\eea
Expansion $\Phi=g\Phi_0+g^2\Phi_1+g^3\Phi_2+\dots$ decomposes this equation
into
\beq
0=\eta_0 Q\Phi_n+\sum_{m=1}^n\sum_{\stackrel{
 \mbox{$\scriptstyle n_1, n_2,\dots, n_{m+1}$}}
{\mbox{$\scriptstyle n_1+n_2+\dots+n_{m+1}=n-m$}}}
 \frac{(-1)^m}{(m+1)!}
 \eta_0[\Phi_{n_1},[\Phi_{n_2},[\dots,[\Phi_{n_m},Q\Phi_{n_{m+1}}]]\dots ].
\eeq
With the condition $b_0\Phi_n=\wt{G}^-_0\Phi_n=0$, we obtain the following
solution.
\beq
\Phi_n=-\frac{\wt{G}^-_0}{L_0}\eta_0\frac{b_0}{L_0}
 \sum_{m=1}^n\sum_{\stackrel{
 \mbox{$\scriptstyle n_1, n_2,\dots, n_{m+1}$}}
{\mbox{$\scriptstyle n_1+n_2+\dots+n_{m+1}=n-m$}}}
 \frac{(-1)^m}{(m+1)!}
 [\Phi_{n_1},[\Phi_{n_2},[\dots,[\Phi_{n_m},Q\Phi_{n_{m+1}}]]\dots ],
\eeq
where
\beq
\wt{G}^-_0=\left[Q,\oint\frac{dz}{2\pi i}zb\xi(z)\right],
\eeq
and we used the following properties:
\beq
\{\eta_0,\wt{G}^-_0\}=L_0,\quad \{Q,\wt{G}^-_0\}=\{b_0,\wt{G}^-_0\}=0.
\eeq
We can show that $\Phi_n$ has properties similar to those of the bosonic case:
$\Phi_n$ consists of states with
$n_--n_+\geq n+1$, where $n_\pm$ are sums of numbers of $\alpha^\pm_{-m}$ and
$\psi_{-r}^\pm$ in the Fock space representation of $\Phi_n$.
Therefore the massless component has no higher
order contribution. Nonzero coefficient of each Fock space state receives
contribution from finite number of $\Phi_n$.
$A_\mu$ is equal to the gauge field of the effective theory $\wh{A}_\mu$.

By the same argument as in \cite{mchsht}, we can also show that
this solution is 1/2 supersymmetric.

%%%%%%%%%%%%%%%%%%%%%%%%%%%%%%%%%%%%%%%%%%%%%%%%%%%%%%%%%%%%%%%%%%%
%%%%%%%%%%%%%%%%%%%%%%%%%%%%%%%%%%%%%%%%%%%%%%%%%%%%%%%%%%%%%%%%%%%
\section{Closed String Field Theory and Field Redefinitions}

In this section we review some basic facts of closed string field theories
and discuss the field redefinitions, which relate components in the string
field and fields in the effective action.

We consider the bosonic closed string field theory\cite{z1,csft} based on
the conformal field theory for the flat spacetime. The string field $\Phi$
is Grassmann even, has ghost number 2, and satisfies $(L_0-\bar{L}_0)\Phi=0$
and $(b_0-\bar{b}_0)\Phi=0$.

First we consider tachyon component of $\Phi$:
\beq
\Phi=\int\frac{d^{26}k}{(2\pi)^{26}}T(k)c\bar{c}e^{ik\cdot X}.
\eeq
$\Phi$ obeys the reality condition that its hermitian conjugate is
minus the BPZ conjugate: ${\rm hc}(\Phi)=-{\rm bpz}(\Phi)$. It gives
$T(k)^\dagger=T(-k)$. Then the quadratic part of the action is
\bea
S^{(2)} & = & -\frac{1}{\ap\kappa^2}\vev{\Phi\Big|c_0^-Q\Big|\Phi} \nn
 & = & \frac{1}{2\kappa^2}\int\frac{d^{26}k}{(2\pi)^{26}}\Bigg[
 -\left(k^2-\frac{4}{\ap}\right)T(-k)T(k)\Bigg].
\eea
This is in the standard form of kinetic term of scalar field with negative
mass squared. Next we consider massless components of $\Phi$:
\bea
\Phi & = & \int\frac{d^{26}k}{(2\pi)^{26}}\Bigg[
\frac{1}{\ap}E_{\mu\nu}(k)c\bar{c}\p X^\mu\bar{\p}X^\nu e^{ik\cdot X}
+E^{(1)}(k) c\p^2 c e^{ik\cdot X}
+E^{(2)}(k) \bar{c}\bar{\p}^2\bar{c}e^{ik\cdot X} \nn
& & +E^{(3)}_\mu(k) (\p c+\bar{\p}\bar{c})c\p X^\mu e^{ik\cdot X}
+E^{(4)}_\mu(k) (\p c+\bar{\p}\bar{c})\bar{c}\bar{\p}X^\mu e^{ik\cdot X}
\Bigg].
\eea
The reality condition gives
\bea
& & E_{\mu\nu}(k)^\dagger=E_{\mu\nu}(-k),\quad
E^{(1)}(k)^\dagger=E^{(1)}(-k),\quad
E^{(2)}(k)^\dagger=E^{(2)}(-k), \nn
& & E^{(3)}_\mu(k)^\dagger=E^{(3)}_\mu(-k),\quad
E^{(4)}_\mu(k)^\dagger=E^{(4)}_\mu(-k).
\eea
The quadratic part of the action is
\bea
S^{(2)} & = & -\frac{1}{\ap\kappa^2}\vev{\Phi|c_0^-Q|\Phi} \nn
 & = & \frac{1}{2\kappa^2}\int\frac{d^{26}k}{(2\pi)^{26}}\Bigg[
 -4\left(E^{(3)}_\mu(-k)-ik_\mu E^{(2)}(-k)
 +\frac{1}{4}ik^\nu E_{\mu\nu}(-k)\right) \nn
 & & \times\left(E^{(3)\mu}(k)+ik^\mu E^{(2)}(k)
 -\frac{1}{4}ik^\lambda E^\mu_{\;\;\lambda}(k)\right) \nn
 & & -4\left(E^{(4)}_\mu(-k)-ik_\mu E^{(1)}(-k)
 -\frac{1}{4}ik^\nu E_{\nu\mu}(-k)\right) \nn
 & & \times\left(E^{(4)\mu}(k)+ik^\mu E^{(1)}(k)
 +\frac{1}{4}ik^\lambda E_\lambda^{\;\;\mu}(k)\right) \nn
 & & +4k_\mu k^\mu\phi(-k)\phi(k)-2k^2\phi(-k)h_\mu^{\;\;\mu}(k)
 +2k_\mu k_\nu\phi(-k)h^{\mu\nu}(k) \nn
 & & +\frac{1}{4}h^{\mu\nu}(-k)\left(-k^2h_{\mu\nu}(k)
 +2k_\mu k_\lambda h_\nu^{\;\;\lambda}(k)
 -2k_\mu k_\nu h_\lambda^{\;\;\lambda}(k)
 +\eta_{\mu\nu}k^2h_\lambda^{\;\;\lambda}(k)\right) \nn
 & & -\frac{1}{12}H_{\mu\nu\lambda}(-k)H^{\mu\nu\lambda}(k)
\Bigg],
\label{mssfta}
\eea
where
\bea
h_{\mu\nu}(k) & = & \Half(E_{\mu\nu}(k)+E_{\nu\mu}(k)),\\
B_{\mu\nu}(k) & = & \Half(E_{\mu\nu}(k)-E_{\nu\mu}(k)),\\
\phi(k) & = & E^{(1)}(k)-E^{(2)}(k)+\frac{1}{4}E_\mu^{\;\;\mu}(k), \\
H_{\mu\nu\lambda}(k) & = & 3ik_{[\mu}B_{\nu\lambda]}(k).
\eea

$E^{(3)}_\mu$ and $E^{(4)}_\mu$ are not dynamical, but are auxiliary fields,
as can be seen from the above quadratic part. We can integrate out them, and
the rest of $S^{(2)}$ coincides with the quadratic part of two derivative
truncation of the low energy effective action $S_{\rm eff}$:
\beq
S_{\rm eff}=\frac{1}{2\wh{\kappa}^2}\int d^{26}x\sqrt{-\wh{g}}
 e^{-2\wh{\phi}}\Bigg[R(\wh{g})
 +4\wh{g}^{\mu\nu}\p_\mu\wh{\phi}\p_\nu\wh{\phi}
 -\frac{1}{12}\wh{H}_{\mu\nu\lambda}\wh{H}^{\mu\nu\lambda}\Bigg],
\eeq
with the following identification\footnote{$\wh{h}_{\mu\nu}=-h_{\mu\nu}$
is also a possible identification, but can be excluded by computing 3-point
interaction term for two tachyons and one graviton.\cite{yz}}:
\bea
\wh{h}_{\mu\nu} & = & h_{\mu\nu},\quad
 \wh{g}_{\mu\nu}=\eta_{\mu\nu}+\wh{h}_{\mu\nu}, \\
\wh{B}_{\mu\nu} & = & B_{\mu\nu},\\
\wh{\phi} & = & \phi+{\rm const}.
\eea
Similarly, if we assume that tachyon field $\wh{T}$ in the effective theory
has the standard form of kinetic term, we can identify it with $T$.

This identification can also be justified by gauge transformation.
Massless part of gauge transformation parameter $\Lambda$ is expanded
as follows:
\bea
\Lambda & = & \int\frac{d^{26}k}{(2\pi)^{26}}\Bigg[
\frac{4}{(\ap)^{3/2}}\epsilon^{(1)}_\mu(k)c\p X^\mu e^{ik\cdot X}
+\frac{4}{(\ap)^{3/2}}\epsilon^{(2)}_\mu(k)\bar{c}\bar{\p}X^\mu e^{ik\cdot X}
 \nn
 & & +\frac{2}{\sqrt{\ap}}\epsilon^{(3)}(k)
(\p c+\bar{\p}\bar{c})e^{ik\cdot X} \Bigg].
\eea
The reality condition ${\rm hc}(\Lambda)=-{\rm bpz}(\Lambda)$ gives
\bea
\epsilon^{(1)}_\mu(k)^\dagger=\epsilon^{(1)}_\mu(-k),\quad
\epsilon^{(2)}_\mu(k)^\dagger=\epsilon^{(2)}_\mu(-k),\quad
\epsilon^{(3)}(k)^\dagger=\epsilon^{(3)}(-k).
\eea
The linearized gauge transformation is
\beq
\delta\Phi=Q\Lambda,
\eeq
which gives
\bea
\delta E_{\mu\nu}(k) & = & 2i(k_\mu\epsilon^{(2)}_\nu(k)
 -k_\nu\epsilon^{(1)}_\mu(k)), \\
\delta E^{(1)}(k) & = & \Half
 ik^\mu\epsilon^{(1)}_\mu(k)+\epsilon^{(3)}(k), \\
\delta E^{(2)}(k) & = & \Half
 ik^\mu\epsilon^{(2)}_\mu(k)+\epsilon^{(3)}(k), \\
\delta E^{(3)}_\mu(k) & = & \Half
 k^2\epsilon^{(1)}_\mu(k)-ik_\mu\epsilon^{(3)}(k), \\
\delta E^{(4)}_\mu(k) & = & \Half
 k^2\epsilon^{(2)}_\mu(k)-ik_\mu\epsilon^{(3)}(k).
\eea
With the following definition of $\epsilon_\mu$ and $\lambda_\mu$,
\bea
\epsilon_\mu & = & \epsilon^{(2)}_\mu-\epsilon^{(1)}_\mu, \\
\lambda_\mu & = & \epsilon^{(1)}_\mu+\epsilon^{(2)}_\mu,
\eea
we obtain
\bea
\delta h_{\mu\nu} & = & ik_\mu\epsilon_\nu+ik_\nu\epsilon_\mu, \\
\delta B_{\mu\nu} & = & ik_\mu\lambda_\nu-ik_\nu\lambda_\mu, \\
\delta\phi & = & 0.
\eea
This is precisely the form of transformation expected from the effective
theory. $\epsilon_\mu$ corresponds to diffeomorphism, and $\lambda_\mu$
corresponds to gauge transformation of B-field. Note that these parameters
do not contain $\epsilon^{(3)}$.

There is no tachyon component in $\Lambda$. Therefore $T$ is invariant
under the linearized gauge transformation. This is the property expected to
$\wh{T}$.

After integrating out $E^{(3)}_\mu$ and $E^{(4)}_\mu$ in (\ref{mssfta}),
$E_{\mu\nu}$, $E^{(1)}$ and $E^{(2)}$ remain. However, $E^{(1)}$ and
$E^{(2)}$ appears only in the combination of $E^{(1)}-E^{(2)}$.
$E^{(1)}-E^{(2)}$ is called ghost dilaton. The other combination
$E^{(1)}+E^{(2)}$ may appear in the interaction terms. There is no field
in the effective theory corresponding to $E^{(1)}+E^{(2)}$. So, if we want,
we can eliminate this field by the gauge transformation:
\beq
\delta(E^{(1)}(k)+E^{(2)}(k))=\Half
 ik^\mu(\epsilon^{(1)}_\mu(k)+\epsilon^{(2)}_\mu(k))+2\epsilon^{(3)}(k)
\eeq
with appropriate choice of $\epsilon^{(3)}$.

The above identification is valid only at the linearized level.
In general, gauge transformation
of $E_{\mu\nu}$ and $\phi$ are different from those of
$\wh{E}_{\mu\nu}\equiv\wh{h}_{\mu\nu}+\wh{B}_{\mu\nu}$ and $\wh{\phi}$
which has direct geometrical and physical meaning. These are related by
``field redefinitions'' after integrating out all the massive components.
$\wh{E}_{\mu\nu}$, $\wh{\phi}$ and $\wh{T}$ are respectively equal to
$E_{\mu\nu}$, $\phi$ and $T$ plus correction terms which consist of two or
more $E_{\mu\nu}$, $\phi$ and $T$, and can have derivatives:
\bea
\wh{E}_{\mu\nu} & = & E_{\mu\nu}
+\mbox{(terms quadratic or higher in $E_{**}$, $\phi$ and $T$)}, 
\label{eredef1} \\
\wh{\phi} & = & {\rm const.}+\phi
+\mbox{(terms quadratic or higher in $E_{**}$, $\phi$ and $T$)},
\label{phiredef1} \\
\wh{T} & = & T
+\mbox{(terms quadratic or higher in $E_{**}$, $\phi$ and $T$)}.
\label{tredef1}
\eea
We take such a normalization of fields that the coupling constant $\kappa$
and $\wh{\kappa}$ appears only as the overall factors of the actions.
Then, as we can see from the procedure described below, the field
redefinitions do not contain them.

Detail of the calculation of the correction terms in (\ref{eredef1}),
(\ref{phiredef1}) and (\ref{tredef1}) is analogous to the
open string case in \cite{cst} (See also \cite{gs}):
We take some appropriate partial gauge fixing condition, for example Siegel
gauge for massive modes and $E^{(1)}+E^{(2)}=0$ for massless modes.
This system still has residual gauge symmetry corresponding to
diffeomorphism and gauge transformation of B-field.
This symmetry is a sum of gauge transformation with the parameters
$\epsilon_\mu$ and $\lambda_\mu$, and the compensating gauge
transformation for maintaining the above partial gauge fixing condition.
The compensating transformation has no linear terms in massless and tachyon
components.
We solve equations of motion for massive modes and massless auxiliary fields.
Then by using them the residual transformation
can be rewritten in terms of $E_{\mu\nu}$, $\phi$ and $T$. With a choice of
covariant gauge fixing condition, which we assume to take, the rewrited
results are covariant under 26 dimensional Lorentz transformation.
If we like, we can also integrate out $T$, regarding it as a kind of massive
mode.

Next we enumerate all the possible terms entering the field redefinitions of
$\wh{E}_{\mu\nu}$, $\wh{\phi}$ and $\wh{T}$ and determine their coefficients
by requiring that $\wh{E}_{\mu\nu}$, $\wh{\phi}$ and $\wh{T}$
have the standard form of the gauge transformation:
\bea
\delta\wh{E}_{\mu\nu} & = & \wh{D}_\mu\wh{\epsilon}_\nu
 +\wh{D}_\nu\wh{\epsilon}_\mu+\p_\mu\wh{\epsilon}^\lambda\wh{B}_{\lambda\nu}
 +\p_\nu\wh{\epsilon}^\lambda\wh{B}_{\mu\lambda}
 -\wh{\epsilon}^\lambda\p_\lambda\wh{B}_{\mu\nu}
 +\p_\mu\wh{\lambda}_\nu-\p_\nu\wh{\lambda}_\mu,\\
\delta\wh{\phi} & = & 0, \\
\delta\wh{T} & = & 0,
\eea
under gauge transformation of $E_{\mu\nu}$, $\phi$ and $T$ plus the
``trivial'' symmmetry\cite{gs}, where $\wh{D}_\mu$ is covariant derivative
with respect to $\wh{g}_{\mu\nu}$. Terms in the field redefinitions are
covariant if we choose a covariant gauge fixing condition. 
Transformation parameters $\wh{\epsilon}_\mu$ and
$\wh{\lambda}_\mu$, for diffeomorphism and gauge transformation of B-field
respectively, are also determined in terms of $\epsilon_\mu$,
$\lambda_\mu$, $E_{\mu\nu}$, $\phi$ and $T$.

Note that the above procedure of integrating out fields is classical.
Therefore we are taking only $\ap$-correction into account. We do not
consider string loop correction. Correspondingly we only consider
$\ap$-correction to $S_{\rm eff}$.

Some ambiguities remain after this procedure.
Firstly we can add gauge covariant terms.
\bea
\wh{E}'_{\mu\nu} & = & \wh{E}_{\mu\nu}
 +U_{\mu\nu}(\wh{g}, \wh{R}, \wh{H}, \wh{\phi}, \wh{T}, \wh{D}_\mu), \\
\wh{\phi}' & = & \wh{\phi}
 +V(\wh{g}, \wh{R}, \wh{H}, \wh{\phi}, \wh{T}, \wh{D}_\mu), \\
\wh{T}' & = & \wh{T}
 +W(\wh{g}, \wh{R}, \wh{H}, \wh{\phi}, \wh{T}, \wh{D}_\mu),
\eea
where $U_{\mu\nu}$, $V$ and $W$ are arbitrary functionals made of
$\wh{g}_{\mu\nu}$, $\wh{\phi}$, $\wh{T}$, $\wh{D}_\mu$, Riemann tensor
$\wh{R}$ with respect to $\wh{g}_{\mu\nu}$, and field strength $\wh{H}$
with respect to $\wh{B}_{\mu\nu}$. Contractions of indices in $U_{\mu\nu}$,
$V$ and $W$ are with $\wh{g}_{\mu\nu}$.

Secondly, we can make gauge transformations:
\bea
\wh{E}'_{\mu\nu}(x) & = &
 \left(\delta_\mu^{\;\;\nu}+\p_\mu\alpha^\nu\right)^{-1\;\lambda}_{\;\;\mu}
 \left(\delta_\mu^{\;\;\nu}+\p_\mu\alpha^\nu\right)^{-1\;\rho}_{\;\;\nu}
 \wh{E}_{\lambda\rho}(x-\alpha), \\
\wh{\phi}'(x) & = & \wh{\phi}(x-\alpha), \\
\wh{T}'(x) & = & \wh{T}(x-\alpha),
\eea
or
\bea
\wh{E}'_{\mu\nu} & = & \wh{E}_{\mu\nu}+\p_\mu\beta_\nu-\p_\nu\beta_\mu, \\
\wh{\phi}' & = & \wh{\phi}, \\
\wh{T}' & = & \wh{T},
\eea
where $\alpha^\mu=\alpha^\mu(\wh{E}, \wh{\phi}, \wh{T}, \p)$ and
$\beta_\mu=\beta_\mu(\wh{E}, \wh{\phi}, \wh{T}, \p)$ are arbitrary
functionals made of $\wh{E}_{\mu\nu}$, $\wh{\phi}$, $\wh{T}$ and $\p_\mu$.

These $\wh{E}'_{\mu\nu}$, $\wh{\phi}'$ and $\wh{T}'$ are equally qualified
as fields
appearing in the effective theory, in terms of their gauge transformations.
These ambiguities can be regarded as field redefinitions within the effective
theory, rather than as relations between the string field theory and
the effective field theory. Note that only terms with derivatives are
involved with the second ambiguity. Terms made of $E_{\mu\nu}$, $\phi$ and $T$
without $\p_\mu$ are not affected by it. In the first ambiguity
terms containing $E_{\mu\nu}$ also have derivatives,
except in terms in the form of $Y(\wh{\phi}, \wh{T}, \wh{D})\wh{g}_{\mu\nu}$
in $U_{\mu\nu}$.

We can fix the first ambiguity for terms linear in $E_{\mu\nu}$, $\phi$ or
$T$ in (\ref{eredef1}), (\ref{phiredef1}) and (\ref{tredef1}): We know that
the linear terms of (\ref{eredef1}), (\ref{phiredef1}) and (\ref{tredef1})
correctly reproduce the quadratic part of the standard
form of the effective action $S_{\rm eff}$. Therefore we cannot add linear
terms any more.

Higher derivative terms in $S_{\rm eff}$ arise in cubic or higher order in
fields. Field redefinitions with higher derivative terms make those terms look
different, and we have no canonical choice for them. Therefore we have no
canonical way to fix the first ambiguity for higher terms.

We can restrict the field redefinitions further by using the
dilaton theorem \cite{bz}: A constant shift of the ghost dilaton is equivalent
to a shift of $\kappa$. In the effective theory a constant shift of
$\wh{\phi}$ is also equivalent to a shift of $\wh{\kappa}$, because we do
not consider string loop correction and
$\wh{\phi}$ should appear with derivatives except the overall factor
$e^{-2\wh{\phi}}$ in the effective action. Therefore we can identify these
two shifts, and correction terms in the field redefinitions should not
contain $\phi$ without derivative. This restricts the first ambiguity further:
Terms with no derivative in the first ambiguity should have $T$,
and therefore terms without $T$ always have derivatives.

In summary the field redefinitions are given by
\bea
\wh{E}_{\mu\nu} & = & E_{\mu\nu}
\nn & & +(\mbox{terms which are quadratic or higher in $E_{**}$, $\p_*\phi$
 and $T$,}
\nn & & \mbox{and may have more derivatives}), \label{eredef2} \\
\wh{\phi}  & = & {\rm const.}+\phi
\nn & & +(\mbox{terms which are quadratic or higher in $E_{**}$, $\p_*\phi$
 and $T$,}
\nn & & \mbox{and may have more derivatives}), \label{phiredef2} \\
\wh{T}  & = & T
\nn & & +(\mbox{terms which are quadratic or higher in $E_{**}$, $\p_*\phi$
 and $T$,}
\nn & & \mbox{and may have more derivatives}),
\eea
and the ambiguities affect terms with derivatives, and terms with $T$.
This is because we have no canonical expression for the effective action of
the tachyon, even for two derivative truncation.

By the procedure described above we can compute the field redefinitions
order by order, but in most cases coefficients can be determined only
numerically. In the next section we will see that some coefficients can be
determined analytically. But before going to the next section we repeat the
above discussion in the heterotic string field theory constructed in
\cite{oz,boz}.

We take superstring CFT as the left mover (without bars), and bosonic CFT as
the right mover (with bars).
In this theory the string field $\Phi$ is Grassmann odd, has ghost number 1
and picture number 0. As in the bosonic case, $(b_0-\bar{b}_0)\Phi=(L_0-
\bar{L}_0)\Phi=0$. Throughout this paper we set R sector components zero.
The massless part of $\Phi$ is
\bea
\Phi & = & \int\frac{d^{10}k}{(2\pi)^{10}}\Bigg[
\frac{i}{\sqrt{2\ap}}E_{\mu\nu}(k)
 \xi c\psi^\mu e^{-\phi}\bar{c}\bar{\p}X^\nu e^{ik\cdot X}
+\frac{\sqrt{\ap}}{2}A^a_\mu(k)
 \xi c\psi^\mu e^{-\phi}\bar{c}\bar{J}^a e^{ik\cdot X}
\nn & &
+E^{(1)}_\mu(k)\bar{c}\bar{\p}X^\mu e^{ik\cdot X}
+iE^{(2)a}(k)\bar{c}\bar{J}^a e^{ik\cdot X}
+E^{(3)}(k)\xi\p\xi ce^{-2\phi}\bar{c}\bar{\p}^2\bar{c}e^{ik\cdot X}
\nn & &
+E^{(4)}(k)\xi\eta ce^{ik\cdot X}
+E^{(5)}_\mu(k)c\p X^\mu e^{ik\cdot X}
+E^{(6)}_{\mu\nu}(k)c\psi^\mu\psi^\nu e^{ik\cdot X}
\nn & &
+iE^{(7)}_\mu(k)\eta e^\phi\psi^\mu e^{ik\cdot X}
+E^{(8)}(k)\p\phi ce^{ik\cdot X}
\nn & &
+iE^{(9)}_\mu(k)(\p c+\bar{\p}\bar{c})\xi c\psi^\mu e^{-\phi} e^{ik\cdot X}
+E^{(10)}(k)(\p c+\bar{\p}\bar{c})e^{ik\cdot X}
\nn & &
+E^{(11)}_\mu(k)(\p c+\bar{\p}\bar{c})
 \xi\p\xi ce^{-2\phi}\bar{c}\bar{\p}X^\mu e^{ik\cdot X}
\nn & &
+iE^{(12)a}(k)(\p c+\bar{\p}\bar{c})
 \xi\p\xi ce^{-2\phi}\bar{c}\bar{J}^a e^{ik\cdot X}
\Bigg],
\eea
where $\bar{J}^a$ is the current of $SO(32)$ or $E_8\times E_8$ which forms
level 1 current algebra:
\beq
\bar{J}^a(z)\bar{J}^b(0)=\frac{\delta^{ab}}{z^2}
+\frac{if_{abc}\bar{J}^c(0)}{z}+({\rm reg.}),
\eeq
where we take such a normalization that the length of roots is 2.
The reality condition ${\rm bpz}(\Phi)=-{\rm hc}(\Phi)$ gives
\bea
& & E_{\mu\nu}(k)^\dagger=E_{\mu\nu}(-k),\quad
A^a_\mu(k)^\dagger=A^a_\mu(-k),\nn
& & E^{(1)}(k)^\dagger=E^{(1)}(-k),\quad
E^{(2)a}(k)^\dagger=E^{(2)a}(-k),\quad
E^{(3)}_\mu(k)^\dagger=E^{(3)}_\mu(-k),\nn
& & E^{(4)}_\mu(k)^\dagger=E^{(4)}_\mu(-k),\quad
E^{(5)}(k)^\dagger=E^{(5)}(-k),\quad
E^{(6)}_{\mu\nu}(k)^\dagger=E^{(6)}_{\mu\nu}(-k),\nn
& & E^{(7)}_\mu(k)^\dagger=E^{(7)}_\mu(-k),\quad
E^{(8)}(k)^\dagger=E^{(8)}(-k),\quad
E^{(9)}_\mu(k)^\dagger=E^{(9)}_\mu(-k),\nn
& & E^{(10)}(k)^\dagger=E^{(10)}(-k),\quad
E^{(11)}_\mu(k)^\dagger=E^{(11)}_\mu(-k),\quad
E^{(12)a}(k)^\dagger=E^{(12)a}(-k).
\eea
The quadratic part of the action is
\bea
S^{(2)} & = & \frac{1}{\ap\kappa^2}\vev{\eta_0\Phi_0\Big|c_0^-Q\Big|\Phi_0}
 \nn
 & = & \frac{1}{2\kappa^2}\int\frac{d^{10}k}{(2\pi)^{10}}\Bigg[
-\frac{8}{\ap}\left(E^{(9)}_\mu(-k)+\sqrt{\frac{\ap}{2}}ik_\mu E^{(3)}(-k)
 -\frac{\sqrt{\ap}}{4\sqrt{2}}ik^\nu E_{\mu\nu}(-k)\right)
\nn & &
 \times\left(E^{(9)\mu}(k)-\sqrt{\frac{\ap}{2}}ik^\mu E^{(3)}(k)
 +\frac{\sqrt{\ap}}{4\sqrt{2}}ik^\lambda E^\mu_{\;\;\lambda}(k)\right)
\nn & &
-4\left(E^{(11)}_\mu(-k)+\Half ik_\mu E^{(4)}(-k)
 +\frac{1}{4}ik^\nu E_{\nu\mu}(-k)\right)
\nn & &
 \times\left(E^{(11)\mu}(k)-\Half ik^\mu E^{(4)}(k)
 -\frac{1}{4}ik^\lambda E_\lambda^{\;\;\mu}(k)\right)
\nn & &
-\frac{16}{\ap}\left(E^{(12)a}(-k)
 -\frac{\ap}{4\sqrt{2}}ik^\mu A^a_\mu(-k)\right)
 \left(E^{(12)a}(k)+\frac{\ap}{4\sqrt{2}}ik^\nu A^a_\nu(k)\right)
\nn & &
 -\frac{\ap}{4}F^a_{\mu\nu}(-k)F^{a\mu\nu}(k)
\nn & &
 +4k^2\phi(-k)\phi(k)-2k^2\phi(-k)h_\mu^{\;\;\mu}(k)
 +2k_\mu k_\nu\phi(-k)h^{\mu\nu}(k)
\nn & &
 +\frac{1}{4}h^{\mu\nu}(-k)\left(-k^2h_{\mu\nu}(k)
 +2k_\mu k_\lambda h_\nu^{\;\;\lambda}(k)
 -2k_\mu k_\nu h_\lambda^{\;\;\lambda}(k)
 +\eta_{\mu\nu}k^2h_\lambda^{\;\;\lambda}(k)\right)
\nn & &
 -\frac{1}{12}H_{\mu\nu\lambda}(-k)H^{\mu\nu\lambda}(k)
\Bigg],
\eea
where
\bea
h_{\mu\nu}(k) & = & \Half(E_{\mu\nu}(k)+E_{\nu\mu}(k)), \\
B_{\mu\nu}(k) & = & \Half(E_{\mu\nu}(k)-E_{\nu\mu}(k)), \\
\phi(k) & = & \Half(E^{(4)}(k)-2E^{(3)}(k))+\frac{1}{4}E_\mu^{\;\;\mu}(k), \\
F^a_{\mu\nu}(k) & = & ik_\mu A^a_\nu(k)-ik_\nu A^a_\mu(k), \\
H_{\mu\nu\lambda}(k) & = & 3ik_{[\mu}B_{\nu\lambda]}(k).
\eea
Note that $E_{**}$ enters $\phi$ in the form of $\frac{1}{4}E_\mu^{\;\;\mu}$
as in the bosonic case, and $E^{(3)}(k)$ and $E^{(4)}(k)$ appear only in 
the combination of $E^{(4)}(k)-2E^{(3)}(k)$.

After integrating out auxiliary fields $E^{(9)}_\mu$, $E^{(11)}_\mu$ and
$E^{(12)}_{AB}$, the above action reproduces the quadratic part of the
two derivative truncation of the effective action $S_{\rm eff}$:
\bea
S_{\rm eff} & = & \frac{1}{2\wh{\kappa}^2}\int d^{10}x\sqrt{-\wh{g}}
 e^{-2\wh{\phi}}\Bigg[R(\wh{g})
 +4\wh{g}^{\mu\nu}\p_\mu\wh{\phi}\p_\nu\wh{\phi}
 -\frac{1}{12}\wh{H}_{\mu\nu\lambda}\wh{H}^{\mu\nu\lambda}
 -\frac{\ap}{4}\wh{F}^a_{\mu\nu}\wh{F}^{a\mu\nu} \nn
 & & +\mbox{(terms with Chern-Simons 3-form)}\Bigg],
\eea
with the following identification:
\bea
\wh{h}_{\mu\nu} & = & h_{\mu\nu}, \label{heth} \\
\wh{B}_{\mu\nu} & = & B_{\mu\nu}, \label{hetb} \\
\wh{\phi} & = & \phi+{\rm const.}, \label{hetphi} \\
\wh{A}^a_\mu & = & A^a_\mu. \label{heta}
\eea

The linearized gauge symmetry is
\beq
\delta\Phi_0=Q\Lambda_0+\eta_0\Lambda_1.
\label{hgtrans}
\eeq
The massless part of $\Lambda_0$ is
\bea
\Lambda_0 & = & \int\frac{d^{10}k}{(2\pi)^{10}}\Bigg[
i\epsilon^{(1)}_\mu(k)\xi c\psi^\mu e^{-\phi}e^{ik\cdot X}
+\epsilon^{(2)}(k)e^{ik\cdot X}
+\epsilon^{(3)}_\mu(k)\xi\p\xi ce^{-2\phi}\bar{c}\bar{\p}X^\mu e^{ik\cdot X}
\nn & &
+i\epsilon^{(4)a}(k)\xi\p\xi ce^{-2\phi}\bar{c}\bar{J}^ae^{ik\cdot X}
+\epsilon^{(5)}(k)(\p c+\bar{\p}\bar{c})\xi\p\xi ce^{-2\phi}e^{ik\cdot X}
\Bigg],
\eea
and the reality condition ${\rm bpz}(\Lambda_0)=+{\rm hc}(\Lambda_0)$ gives
\bea
& & \epsilon^{(1)}_\mu(k)^\dagger=\epsilon^{(1)}_\mu(-k),\quad
\epsilon^{(2)}(k)^\dagger=\epsilon^{(2)}(-k),\quad
\epsilon^{(3)}_\mu(k)^\dagger=\epsilon^{(3)}_\mu(-k),\nn
& & \epsilon^{(4)a}(k)^\dagger=\epsilon^{(4)a}(-k),\quad
\epsilon^{(5)}(k)^\dagger=\epsilon^{(5)}(-k).
\eea
The first term of (\ref{hgtrans}) gives
\bea
\delta E_{\mu\nu} & = & i\sqrt{2\ap}k_\nu\epsilon^{(1)}_\mu(k)
 -\ap ik_\mu\epsilon^{(3)}_\nu(k), \\
\delta A^a_\mu & = & \sqrt{2}ik_\mu\epsilon^{(4)a}(k), \\
\delta E^{(1)}_\mu & = & ik_\mu\epsilon^{(2)}(k)-\epsilon^{(3)}_\mu(k), \\
\delta E^{(2)a} & = & -\epsilon^{(4)a}(k), \\
\delta E^{(3)} & = & -\frac{\ap}{4}ik^\mu\epsilon^{(3)}_\mu(k)
 +\epsilon^{(5)}(k), \\
\delta E^{(4)} & = & -\sqrt{\frac{\ap}{2}}ik^\mu\epsilon^{(1)}_\mu(k)
 +2\epsilon^{(5)}(k), \\
\delta E^{(5)}_\mu & = & -\sqrt{\frac{2}{\ap}}\epsilon^{(1)}_\mu(k)
 +ik_\mu\epsilon^{(2)}(k), \\
\delta E^{(6)}_{\mu\nu} & = &
 \Half\sqrt{\frac{\ap}{2}}(ik_\nu\epsilon^{(1)}_\mu(k)
 -ik_\mu\epsilon^{(1)}_\nu(k)), \\
\delta E^{(7)}_\mu & = & \epsilon^{(1)}_\mu(k)
 -\sqrt{\frac{\ap}{2}}ik_\mu\epsilon^{(2)}(k), \\
\delta E^{(8)} & = & \sqrt{\frac{\ap}{2}}ik^\mu\epsilon^{(1)}_\mu(k)
 -2\epsilon^{(5)}(k), \\
\delta E^{(9)}_\mu & = & \frac{\ap}{4}k^2\epsilon^{(1)}_\mu(k)
 +\sqrt{\frac{\ap}{2}}ik_\mu\epsilon^{(5)}(k), \\
\delta E^{(10)} & = & \frac{\ap}{4}k^2\epsilon^{(2)}(k)
 +\epsilon^{(5)}(k), \\
\delta E^{(11)}_\mu & = & \frac{\ap}{4}k^2\epsilon^{(3)}_\mu(k)
 +ik_\mu\epsilon^{(5)}(k), \\
\delta E^{(12)a} & = & \frac{\ap}{4}k^2\epsilon^{(4)a}(k).
\eea
With the following definition of $\epsilon_\mu$, $\lambda_\mu$, $\omega^a$,
\bea
\epsilon_\mu(k) & = & \sqrt{\frac{2}{\ap}}\epsilon^{(1)}_\mu(k)
 -\frac{2}{\ap}\epsilon^{(3)}_\mu(k), \\
\lambda_\mu(k) & = & -\sqrt{\frac{2}{\ap}}\epsilon^{(1)}_\mu(k)
 -\frac{2}{\ap}\epsilon^{(3)}_\mu(k), \\
\omega^a(k) & = & \sqrt{2}\epsilon^{(4)a}(k),
\eea
the above transformation reproduces precisely the expected form:
\bea
\delta h_{\mu\nu}(k) & = & ik_\mu\epsilon_\nu(k)+ik_\nu\epsilon_\mu(k), \\
\delta B_{\mu\nu}(k) & = & ik_\mu\lambda_\nu(k)-ik_\nu\lambda_\mu(k), \\
\delta\phi(k) & = & 0, \\
\delta A^a_\mu(k) & = & ik_\mu\omega^a(k).
\eea
Note that this does not contain $\epsilon^{(2)}$ and $\epsilon^{(5)}$.
$\epsilon^{(2)}$ enters only those without $\xi_0$. 
$\Lambda_1$ in the second term of (\ref{hgtrans}) does not enter
$h_{\mu\nu}$, $B_{\mu\nu}$, $\phi$ and $A^a_\mu$.

$E^{(4)}+2E^{(3)}$ and those without $\xi_0$ do not appear in the quadratic
action, and may appear in the interaction terms.
These can be gauged away by $\Lambda_1$ and $\epsilon^{(5)}$.

We can expect that the dilaton theorem also holds in this theory:
A shift of $E^{(4)}-2E^{(3)}$ is equivalent to a shift of $\kappa$.
Although we have no proof, there is evidence for it\cite{okw}.
Therefore we assume it.

(\ref{heth}), (\ref{hetb}), (\ref{hetphi}) and (\ref{heta}) are correct only
in the linearized case, and have correction terms.
Procedure for determining them is analogous to the bosonic case.
When $A^a_\mu=0$, those are given by (\ref{eredef2}) and (\ref{phiredef2})
(with $T=0$).

%%%%%%%%%%%%%%%%%%%%%%%%%%%%%%%%%%%%%%%%%%%%%%%%%%%%%%%%%%%%%%%%%%%
%%%%%%%%%%%%%%%%%%%%%%%%%%%%%%%%%%%%%%%%%%%%%%%%%%%%%%%%%%%%%%%%%%%
\section{Field redefinitions and T-duality}

In this section we discuss restriction on the field redefinitions imposed by
T-duality transformation. For an early study of T-duality in string field
theory see \cite{kz}.

We divide spacetime coordinates $x^\mu$ into $x^i$ and $x^a$.
$x^i$ are directions which T-duality transformation is applied to,
and $x^a$ are the rest. \footnote{Indices $a$, $b$, $\dots$ in this section
should not be confused with
those for the gauge group $SO(32)$ or $E_8\times E_8$ in the heterotic case.}
We assume that $x^i$ are compactified to a rectangular torus.

In the conformal field theory in the flat spacetime T-duality transformation
is identified with the parity transformation for the right moving sector:
\beq
X^i_R(\bar{z})\rightarrow -X^i_R(\bar{z}),\quad
X^a_R(\bar{z})\rightarrow X^a_R(\bar{z}).
\eeq
This transformation is extended to string field theory with no modification,
and relates two string field theories in two different tori.
In terms of coefficient fields in the string field it is expressed by
exchanging momentum and winding modes, and putting minus signs on
component fields if odd number of indices are contracted to $X^i_R$ in the
string field.

In this section we consider only the sector which has no momentum and
no winding number along $x^i$, and T-dualized quantities are denoted by
primed symbols. Then the above transformation transforms massless modes
in the string field as follows:
\bea
E'_{\mu a} & = & E_{\mu a}, \label{tema} \\
E'_{\mu i} & = & -E_{\mu i}, \label{temi} \\
\phi' & = & \phi-\Half E_{ii}. \label{tphi}
\eea
On the effective theory side we have well-known T-duality rule\cite{bsch}:
\bea
\wh{E}'_{ij} & = & -\delta_{ij}+(\delta_{ij}+\wh{E}_{ij})^{-1}_{ij},
\label{bscheij} \\
\wh{E}'_{aj} & = & -\wh{E}_{ak}(\delta_{ij}+\wh{E}_{ij})^{-1}_{kj},
\label{bscheaj} \\
\wh{E}'_{ib} & = & (\delta_{ij}+\wh{E}_{ij})^{-1}_{ik}\wh{E}_{kb},
\label{bscheib} \\
\wh{E}'_{ab} & = & \wh{E}_{ab}
 -\wh{E}_{ai}(\delta_{ij}+\wh{E}_{ij})^{-1}_{ij} \wh{E}_{jb},
\label{bscheab} \\
e^{2\wh{\phi}'} & = & e^{2\wh{\phi}} {\rm det}(\delta_{ij}+\wh{E}_{ij})^{-1},
\label{bschphi}
\eea
where we consider only the case where $T=0$ in bosonic theory and $A^a_\mu=0$
in heterotic theory.\footnote{
In some literature the signs of the right hand sides of (\ref{bscheaj}) and
(\ref{bscheib}) are opposite. This corresponds to flipping sign of B-field,
or exchanging the role of the left mover and the right mover.}
This is consistent because $T$ and $\wh{T}$ are invariant
under T-duality transformation, and $A^a_\mu=0$ implies
$A^{'a}_\mu=\wh{A}^a_\mu=\wh{A}^{'a}_\mu=0$.
Note that the above T-duality rule is the lowest order relation in $\ap$
and receives higher derivative corrections.

When linearized, (\ref{tema}), (\ref{temi}) and (\ref{tphi})
coincide with (\ref{bscheij}), (\ref{bscheaj}), (\ref{bscheib}),
(\ref{bscheab}) and (\ref{bschphi}). Therefore it is natural to
identify those transformations.
Then we can give some information on the nonderivative part of the
relation between $E_{\mu\nu}$ and $\wh{E}_{\mu\nu}$:
Since we consider only zero momentum and zero winding sector,
the field redefinitions in the string field theories on the both tori are
in the same form. Therefore if we write down general form of the field
redefinitions and plug them into (\ref{bscheij})-(\ref{bschphi}), then we can
determine coefficients. Because (\ref{bscheij})-(\ref{bschphi}) are lowest
order relation and we can give no information on terms with derivatives,
in this section we neglect those terms.

Let us note that the following $e_{\mu\nu}$
\beq
e_{\mu\nu}=E_{\mu\nu}+\sum_{n=1}^\infty\frac{1}{2^n}E_\mu^{\;\;\lambda_1}
 E_{\lambda_1}^{\;\;\lambda_2}\dots E_{\lambda_n\nu}
=\left(\delta_\mu^{\;\;\lambda}
 -\Half E_\mu^{\;\;\lambda}\right)_\mu^{-1\lambda}E_{\lambda\nu}
\eeq
satisfies (\ref{bscheij})-(\ref{bscheab}) if we set
$\wh{E}_{\mu\nu}=e_{\mu\nu}$. This can be proven as follows. $e_{\mu\nu}$
satisfies
\beq
e_{\mu\nu}=E_{\mu\nu}+\Half E_\mu^{\;\;\lambda}e_{\lambda\nu}
 =E_{\mu\nu}+\Half e_\mu^{\;\;\lambda}E_{\lambda\nu}.
\label{eiden}
\eeq
First let us show (\ref{bscheij}), or equivalently
$e_{ij}+e_{ij}'+e_{ik}e'_{kj}=0$. From (\ref{eiden}),
\bea
e_{ij} & = & E_{ij}+\Half E_i^{\;\; k}e_{kj}+\Half E_i^{\;\; b}e_{bj},
\label{eij1} \\
e_{aj} & = & E_{aj}+\Half E_a^{\;\; k}e_{kj}+\Half E_a^{\;\; b}e_{bj}.
\label{eaj1}
\eea
From these we can express $e_{ij}$ in terms of $E_{**}$:
\bea
e_{ij} & = & \left(\delta_i^{\;\; k}-\Half E_i^{\;\; k}
 -\frac{1}{4}E_i^{\;\; a}\left(\delta_a^{\;\; b}-\Half E_a^{\;\; b}
 \right)_{\;\; a}^{-1\; b}E_b^{\;\; k}
 \right)^{-1}_{ik} \nn
 & & \times\left(E_{kj}+\Half E_k^{\;\; c}
 \left(\delta_c^{\;\; d}
 -\Half E_c^{\;\; d}\right)_{\;\; c}^{-1\; d}E_{dj}\right).
\label{eij2}
\eea 
Therefore,
\bea
e_{ij}' & = & -\left(\delta_i^{\;\; k}+\Half E_i^{\;\; k}
 +\frac{1}{4}E_i^{\;\; a}\left(\delta_a^{\;\; b}-\Half E_a^{\;\; b}
 \right)_{\;\; a}^{-1\; b}E_b^{\;\; k}
 \right)^{-1}_{ik} \nn
 & & \times\left(E_{kj}+\Half E_k^{\;\; c}
 \left(\delta_c^{\;\; d}
 -\Half E_c^{\;\; d}\right)_{\;\; c}^{-1\; d}E_{dj}\right).
\label{eijp2}
\eea
Note that matrices in the first lines
and second lines of (\ref{eij2}) and (\ref{eijp2}) commute,
because they consist of the same matrix 
$E_{ij}+\Half E_i^{\;\; a}
 \left(\delta_a^{\;\; b}-\Half E_a^{\;\; b}\right)_{\;\; a}^{-1\; b}E_{bj}$.

By using these, it is straightforward calculation to show
\bea
0 & = & \left(\delta_i^{\;\; k}-\Half E_i^{\;\; k}
 -\frac{1}{4}E_i^{\;\; a}\left(\delta_a^{\;\; b}-\Half E_a^{\;\; b}
 \right)_{\;\; a}^{-1\; b}E_b^{\;\; k}\right)
 (e_{kl}+e_{kl}'+e_{km}e_{ml}') \nn
& & \times\left(\delta^{lj}+\Half E^{lj}
 +\frac{1}{4}E^{lc}\left(\delta_c^{\;\; d}-\Half E_c^{\;\; d}
 \right)_{\;\; c}^{-1\; d}E_d^{\;\; j}\right).
\eea
Hence $e_{ij}$ satisfies $e_{ij}+e_{ij}'+e_{ik}e_{kj}'=0$.

Next we show (\ref{bscheaj}) and (\ref{bscheib}). From (\ref{eaj1}),
\beq
e_{aj}=\left(\delta_a^{\;\; b}-\Half E_a^{\;\; b}\right)_{\;\; a}^{-1\; b}
 \left(E_{bj}+\Half E_b^{\;\; k}e_{kj}\right).
\eeq
Similarly,
\beq
e_{ib}=\left(E_i^{\;\; a}+\Half e_i^{\;\; k}E_k^{\;\; a}\right)
 \left(\delta_a^{\;\; b}-\Half E_a^{\;\; b}\right)_{\;\; ab}^{-1}.
\eeq
By using these and (\ref{bscheij}), we can show (\ref{bscheaj}) and
(\ref{bscheib}) as follows.
\bea
e_{aj}' & = & \left(\delta_a^{\;\; b}-\Half E_a^{\;\; b}
 \right)_{\;\; a}^{-1\; b}
 \left(-E_{bj}-\Half E_b^{\;\; k}(-\delta_{kj}
 +(\delta_k^{\;\; j}+e_k^{\;\; j})^{-1}_{\;\; kj})\right) \nn
 & = & -\left(\delta_a^{\;\; b}-\Half E_a^{\;\; b}
 \right)_{\;\; a}^{-1\; b}
 \left(E_b^{\;\; l}+\Half E_b^{\;\; k}e_k^{\;\; l}\right)
 (\delta_l^{\;\; j}+e_l^{\;\; j})^{-1}_{\;\; lj} \nn
 & = & -e_a^{\;\; k}(\delta_k^{\;\; j}+e_k^{\;\; j})^{-1}_{\;\; kj}, \\
e_{ib}' & = & \left(E_i^{\;\; a}
 +\Half(-\delta_i^{\;\; k}
 +(\delta_i^{\;\; k}+e_i^{\;\; k})^{-1\; k}_{\;\; i})E_k^{\;\; a}\right)
 \left(\delta_a^{\;\; b}-\Half E_a^{\;\; b}\right)_{\;\; ab}^{-1} \nn
 & = & (\delta_i^{\;\; j}+e_i^{\;\; j})^{-1}_{\;\; ij}
 \left(E^{ja}+\Half e^j_{\;\; k}E^{ka}\right)
 \left(\delta_a^{\;\; b}-\Half E_a^{\;\; b}\right)_{\;\; ab}^{-1} \nn
 & = & (\delta_i^{\;\; k}+e_i^{\;\; k})^{-1\; k}_{\;\; i}e_{kb}.
\eea

Finally, from (\ref{eiden}),
\beq
e_{ab}=\left(\delta_a^{\;\; c}-\Half E_a^{\;\; c}\right)_{\;\; a}^{-1\; c}
 \left(E_{cb}+\Half E_c^{\;\; i}e_{ib}\right).
\eeq
Therefore, from (\ref{bscheib}),
\bea
e_{ab}' & = & \left(\delta_a^{\;\; c}
 -\Half E_a^{\;\; c}\right)_{\;\; a}^{-1\; c}
 \left(E_{cb}-\Half E_c^{\;\; i}
 (\delta_i^{\;\; j}+e_i^{\;\; j})^{-1\; j}_{\;\; i}e_{jb}\right) \nn
 & = & e_{ab}-\Half\left(\delta_a^{\;\; c}
 -\Half E_a^{\;\; c}\right)_{\;\; a}^{-1\; c}
 E_c^{\;\; i}(\delta_i^{\;\; j}
 +(\delta_i^{\;\; j}+e_i^{\;\; j})^{-1\; j}_{\;\; i})e_{jb} \nn
 & = & e_{ab}-e_a^{\;\; i}
 (\delta_i^{\;\; j}+e_i^{\;\; j})^{-1\; j}_{\;\; i}e_{jb}.
\eea

Thus we have shown that $e_{\mu\nu}$ satisfies (\ref{bscheij})-(\ref{bscheab})
and is a good candidate for $\wh{E}_{\mu\nu}$. Is this the general solution for
(\ref{bscheij})-(\ref{bscheab})? The following is a partial answer to it,
which is the main result of this section:
\bea
\wh{E}_{\mu\nu} & = & e_{\mu\nu} \nn
 & & +\mbox{(terms in which $\mu$ is a first index of $E_{**}$ and
 $\nu$ is a second index of  $E_{**}$,} \nn
 & & \mbox{and which contain at least one 1-1 contraction} \nn
 & & \mbox{and at least one 2-2 contraction)},
\label{eresult} \\
\wh{\phi} & = & c+\phi-\frac{1}{4}E_\mu^{\;\;\mu}
 -\frac{1}{2}{\rm ln}\;{\rm det}
 \left(\delta_\mu^{\;\;\nu}-\Half E_\mu^{\;\;\nu}\right) \nn
 & & +\mbox{(terms which contain at least one 1-1 contraction} \nn
 & & +\mbox{and at least one 2-2 contraction)},
\label{phiresult}
\eea
where $c$ is a constant, and 1-1, 2-1 and 2-2 contractions are defined as
those of type $E_{\lambda *}E^\lambda_{\;\; *}$,
$E_{*\lambda}E^\lambda_{\;\; *}$ and $E_{*\lambda}E_*^{\;\;\lambda}$
respectively. 2-1 contraction can be regarded as matrix product.
Therefore, noting that
\beq
{\rm ln}\;{\rm det}\left(\delta_\mu^{\;\;\nu}-\Half E_\mu^{\;\;\nu}\right)
=-\sum_{n=1}^\infty\frac{1}{2^nn}E_{\lambda_1}^{\;\;\lambda_2}
 E_{\lambda_2}^{\;\;\lambda_3}\dots E_{\lambda_n}^{\;\;\lambda_1},
\eeq
(\ref{eresult}) and (\ref{phiresult}) mean that 
terms written using only matrix product and trace are determined by the
first lines of them, and here we do not attempt to determine coefficients of
terms with 1-1 and 2-2 contractions. Those undetermined terms consist of
only $E_{**}$, and quadratic or higher in it. In (\ref{eresult}), $\mu$ and
$\nu$ may appear in the same $E_{**}$: $E_{\mu\nu}$, or in different $E_{**}$:
$E_{\mu *}E_{*\nu}$.
As we will see later, some of the undetermined coefficients are related to
each other, and some are arbitrary.

To prove (\ref{eresult}), first we show that
$\wh{E}_{\mu\nu}$ does not have the following types of term
by induction on the order of $E_{**}$:
\begin{description}
\item (i) $\eta_{\mu\nu}\times\mbox{(scalar)}$
\item (ii) terms in which $\mu$ is a second index of $E_{**}$ and
 $\nu$ is a second index of $E_{**}$
\item (iii) terms in which $\mu$ is a first index of $E_{**}$ and
 $\nu$ is a first index of $E_{**}$
\item (iv) terms in which $\mu$ is a second index of $E_{**}$ and
 $\nu$ is a first index of $E_{**}$
\end{description}
This is obvious at the linearized order.
Suppose this is correct up to the order $(E_{**})^{n-1}$.
Then from (\ref{bscheij}),
\bea
\wh{E}_{ij}' |_{(E_{**})^n} & = & (-\wh{E}_{ij}+\wh{E}_{ik}\wh{E}_{kj}
 -\wh{E}_{ik}\wh{E}_{kl}\wh{E}_{lj}+\dots)|_{(E_{**})^n} \nn
& = & -\wh{E}_{ij}|_{(E_{**})^n} \nn
& & +\mbox{(terms in which $i$ is a first index of $E_{**}$} \nn
& & \mbox{and $j$ is a second index of  $E_{**}$)}.
\eea
Therefore terms listed above, of the order $(E_{**})^n$,
are in the left hand side and the first term of the right hand side.
This means that those terms must change sign under T-duality transformation.
It is easy to see that terms of type (i), (ii) and (iii) do not satisfy this
requirement. For example, for (i), contractions of
$E_{\lambda *}E^\lambda_{\;\; *}$ and $E_{*\lambda}E_*^{\;\;\lambda}$
do not have sign change from $\lambda$ under T-duality transformation. 
On the other hand, part of
$E_{*\lambda}E^\lambda_{\;\; *}=E_{*a}E^a_{\;\; *}+E_{*i}E^i_{\;\; *}$
have sign change: $E_{*a}E^a_{\;\; *}-E_{*i}E^i_{\;\; *}$. Therefore there
is no scalar made of $E_{**}$ which changes sign under T-duality
transformation. So there is no term of type (i) at this order.
Terms of type (iv) may satisfy the requirement, but from (\ref{bscheaj}),
\bea
\wh{E}_{aj}' |_{(E_{**})^n} & = & -\wh{E}_{aj}|_{(E_{**})^n} \nn
& & +\mbox{(terms in which $a$ is a first index of $E_{**}$} \nn
& & \mbox{and $j$ is a second index of  $E_{**}$)}
\eea
and again order $(E_{**})^n$ terms of type (iv) are
in the left hand side and the first term of the right hand side.
They must change sign under T-duality transformation, and it can be easily
shown that there is no such term.

Thus we have shown that there is no term of type (i)-(iv).
Next we show that $e_{\mu\nu}$ exhausts terms in which $\mu$ is a first
index of $E_{**}$ and $\nu$ is a second index of $E_{**}$,
and there are no 1-1 and 2-2 contractions i.e. there is no more term
of the form $E_\mu^{\;\;\lambda_1}E_{\lambda_1}^{\;\;\lambda_2}\dots
E_{\lambda_n\nu}$.
Note that proving this completes the proof of (\ref{eresult}) because 
if there is at least one 1-1 (or 2-2) contraction, then there is at least
one 2-2 (or 1-1) contraction, as the number of the first indices
and the second indices of $E_{**}$ are equal.

Our proof of this fact is given again by induction.
From (\ref{bscheij}),
\bea
\wh{E}_{ij}' |_{(E_{**})^n} & = & (-\wh{E}_{ij}+\wh{E}_{ik}\wh{E}_{kj}
 -\wh{E}_{ik}\wh{E}_{kl}\wh{E}_{lj}+\dots)|_{(E_{**})^n} \nn
& = & -\wh{E}_{ij}|_{(E_{**})^n} \nn
& & +\mbox{(terms with no 1-1 and 2-2 contractions)} \nn
& & +\mbox{(terms with at least one 1-1 contraction} \nn
& & +\mbox{and at least one 2-2 contraction)}.
\label{ehatij1}
\eea
By the assumption of the induction terms in the second line of the last
expression of (\ref{ehatij1}) are those given by replacing $\wh{E}_{**}$ by
$e_{**}$.
We know that, as we showed earlier, terms coming from only $e_{**}$ cancel
between the left and the right hand sides, and if there is more order
$(E_{**})^n$ term of the form
$E_\mu^{\;\;\lambda_1}E_{\lambda_1}^{\;\;\lambda_2}\dots E_{\lambda_n\nu}$
it must be in the left hand side and the first term of the right hand side.
This means that it must change sign under T-duality transformation.
But it does not. This completes our proof of (\ref{eresult}).

Next we show (\ref{phiresult}). From (\ref{bschphi}),
\beq
\wh{\phi}'-\frac{1}{4}{\rm ln}\;{\rm det}(\delta_{ij}+\wh{E}_{ij}')=
\wh{\phi}-\frac{1}{4}{\rm ln}\;{\rm det}(\delta_{ij}+\wh{E}_{ij})
\label{tinv}
\eeq
i.e. 
\beq
\wh{\phi}=\frac{1}{4}{\rm ln}\;{\rm det}(\delta_{ij}+\wh{E}_{ij})
+\mbox{(terms invariant under T-duality transformation)}.
\eeq
Since $\wh{\phi}$ should consist of terms covariant under 26 (or 10)
dimensional Lorentz transformation, and
${\rm ln}\;{\rm det}(\delta_{ij}+\wh{E}_{ij})$
is not covariant, it should be possible to covariantize 
${\rm ln}\;{\rm det}(\delta_{ij}+\wh{E}_{ij})$ by adding T-duality
invariant terms. Naive covariantization of
${\rm ln}\;{\rm det}(\delta_{ij}+\wh{E}_{ij})$ is
${\rm ln}\;{\rm det}(\delta_\mu^{\;\;\nu}+\wh{E}_\mu^{\;\;\nu})$.
However, since ${\rm ln}\;{\rm det}(\delta_{ij}+\wh{E}_{ij})=
{\rm ln}\;{\rm det}(\delta_\mu^{\;\;\nu}+\wh{E}_\mu^{\;\;\nu})
-{\rm ln}\;{\rm det}(\delta_a^{\;\; b}+\wh{E}_a^{\prime\; b})$
and ${\rm ln}\;{\rm det}(\delta_a^{\;\; b}+\wh{E}_a^{\prime\; b})$
is not T-duality invariant, this cannot be the correct covariantization.

This means that the covariantization is possible only when $\wh{E}_{\mu\nu}$
takes some particular form. This may give a restriction on
possible form of the field redefinitions.
We show that $\wh{E}_{\mu\nu}=e_{\mu\nu}$
allows us to covariantize ${\rm det}(\delta_{ij}+\wh{E}_{ij})$.
From (\ref{eij2}),
\bea
{\rm det}(\delta_{ij}+e_{ij}) & = & {\rm det}\left(\delta_{ij}+\Half E_{ij}
 +\frac{1}{4}E_i^{\;\; a}\left(\delta_a^{\;\; b}
 -\Half E_a^{\;\; b}\right)_{\;\; a}^{-1\; b}E_{bj}\right) \nn
 & & \times{\rm det}\left(\delta_{ij}-\Half E_{ij}
 -\frac{1}{4}E_i^{\;\; a}\left(\delta_a^{\;\; b}
 -\Half E_a^{\;\; b}\right)_{\;\; a}^{-1\; b}E_{bj}\right)^{-1},
\eea
and by straightforward calculation,
\bea
{\rm det}\left(\delta_\mu^{\;\;\nu}-\Half E_\mu^{\;\;\nu}\right) & = &
{\rm det}\left(\delta_a^{\;\; b}-\Half E_a^{\;\; b}\right) \nn
 & & \times{\rm det}\left(\delta_{ij}-\Half E_{ij}
 -\frac{1}{4}E_i^{\;\; a}\left(\delta_a^{\;\; b}
 -\Half E_a^{\;\; b}\right)_{\;\; a}^{-1\; b}E_{bj}\right).
\eea
From these,
\beq
{\rm ln}\;{\rm det}(\delta_{ij}+e_{ij}')
+2\;{\rm ln}\;{\rm det}\left(\delta_\mu^{\;\;\nu}
-\Half E_\mu^{\prime\;\nu}\right)=
{\rm ln}\;{\rm det}(\delta_{ij}+e_{ij})
+2\;{\rm ln}\;{\rm det}\left(\delta_\mu^{\;\;\nu}-\Half E_\mu^{\;\;\nu}\right).
\eeq
By adding 1/4 of the above equation to (\ref{tinv}), we see that
when $\wh{E}_{\mu\nu}=e_{\mu\nu}$,
\bea
\wh{\phi} & = & -\Half{\rm ln}\;{\rm det}
\left(\delta_\mu^{\;\;\nu}-\Half E_\mu^{\;\;\nu}\right) \nn
 & & +\mbox{(terms invariant under T-duality transformation)}.
\eea
We know that $\wh{E}_{\mu\nu}$ may have more terms other than $e_{\mu\nu}$.
In that case, we separate $\frac{1}{4}{\rm ln}\;{\rm det}
\left(\delta_{ij}+\wh{E}_{ij}\right)$ into contributions from $e_{\mu\nu}$
and the rest:
\bea
\frac{1}{4}{\rm ln}\;{\rm det}\left(\delta_{ij}+\wh{E}_{ij}\right)
 & = & \frac{1}{4}{\rm ln}\;{\rm det}
 \left(\delta_{ij}+e_{ij}+(\wh{E}_{ij}-e_{ij})\right) \nn
 & = & \frac{1}{4}{\rm ln}\;{\rm det}(\delta_{ij}+e_{ij})
 +\frac{1}{4}{\rm ln}\;{\rm det}
 \left(\delta_{ij}+(\delta_{ij}+e_{ij})^{-1\; k}_{\;\; i}
 (\wh{E}_{kj}-e_{kj})\right).
\eea
We have just shown that the first term of the above equation can be
covariantized to be $-\Half{\rm ln}\;{\rm det}
\left(\delta_\mu^{\;\;\nu}-\Half E_\mu^{\;\;\nu}\right)$, and
the second term must also be able to covariantized by adding noncovariant 
T-duality invariant terms.
Such covariantized terms have both of 1-1 and 2-2 contractions because
$\wh{E}_{kj}-e_{kj}$ does.
In addition, in $\wh{\phi}$ we may have more
terms which are covariant and T-duality invariant.
$\phi-\frac{1}{4}E_\mu^{\;\;\mu}$ is the only such term linear in
fields, and a constant is also allowed. Quadratic or higher terms are made
of only $E_{**}$, and in order to be T-duality invariant all the
contractions are those of 1-1 or 2-2. This completes our proof of
(\ref{phiresult}).

To see how terms with 1-1 and 2-2 contractions behave, let us determine 
$\wh{E}_{\mu\nu}$ and $\wh{\phi}$ order by order. We start from
$\wh{E}_{\mu\nu}|_{(E_{**})^1}=E_{\mu\nu}$ and $\wh{\phi}|_{(E_{**})^1}=\phi$,
and the next order contribution is determined as follows. From (\ref{bscheij}),
\bea
\wh{E}'_{ij}|_{(E_{**})^2} & = &
 (-\wh{E}_{ij}+\wh{E}_{ik}\wh{E}_{kj})|_{(E_{**})^2} \nn
 & = & -\wh{E}_{ij}|_{(E_{**})^2}+E_{ik}E_{kj} \nn
 & = & -\wh{E}_{ij}|_{(E_{**})^2}
 +\Half(E_i^{\prime\;\mu}E_{\mu j}'+E_i^{\;\;\mu}E_{\mu j}).
\eea
Therefore $\wh{E}_{\mu\nu}|_{(E_{**})^2}
=\Half E_\mu^{\;\;\lambda}E_{\lambda\nu}=e_{\mu\nu}|_{(E_{**})^2}$.
At this order this satisfies (\ref{bscheij})-(\ref{bscheab}), and
\bea
\frac{1}{4}{\rm ln}\;{\rm det}(\delta_{ij}+\wh{E}_{ij}
 )\Bigg|_{(E_{**})^2} & = &
\frac{1}{4}\left(\wh{E}_{ii}-\Half\wh{E}_{ij}\wh{E}_{ji}
 \right)\Bigg|_{(E_{**})^2} \nn
& = & \frac{1}{4}
 \left(\Half E_i^{\;\;\lambda}E_{\lambda i}-\Half E_{ij}E_{ji}\right) \nn
& = & \frac{1}{16}
 \left(E_{\lambda\rho}E^{\rho\lambda}-E_{ij}E_{ji}-E_{ab}E_{ba}\right).
\eea
By adding T-duality invariant term
$\frac{1}{16}\left(E_{ij}E_{ji}+E_{ab}E_{ba}\right)$,
we can covariantize
$-\frac{1}{4}{\rm ln}\;{\rm det}(\delta_{ij}+\wh{E}_{ij})$.
Then at this order $\wh{\phi}$ is given by
\bea
\wh{\phi}|_{(E_{**})^2} & = & \frac{1}{16}E_{\mu\nu}E^{\nu\mu}
 +c^{(2)}_1E_{\mu\nu}E^{\mu\nu} \nn
 & = & -\Half{\rm ln}\;{\rm det}
 \left(\delta_\mu^{\;\;\nu}-\Half E_\mu^{\;\;\nu}\right)\Bigg|_{(E_{**})^2}
 +c^{(2)}_1E_{\mu\nu}E^{\mu\nu},
\eea
where $c^{(2)}_1$ is an arbitrary constant.

Next we investigate order $(E_{**})^3$.
\bea
\wh{E}'_{ij}|_{(E_{**})^3} & = &
 (-\wh{E}_{ij}+\wh{E}_{ik}\wh{E}_{kj}
 -\wh{E}_{ik}\wh{E}_{kl}\wh{E}_{lj})|_{(E_{**})^3} \nn
 & = & -\wh{E}_{ij}|_{(E_{**})^3}
 +\Half E_i^{\;\;\lambda}E_\lambda^{\;\; k}E_{kj}
 +\Half E_i^{\;\; k}E_k^{\;\;\lambda}E_{\lambda j}
 -E_{ik}E_{kl}E_{lj} \nn
 & = & -\wh{E}_{ij}|_{(E_{**})^3}
 +\frac{1}{4}(E_i^{\prime\;\mu}E_\mu^{\prime\;\nu}E_{\nu j}'
 +E_i^{\;\;\mu}E_\mu^{\;\;\nu}E_{\nu j}).
\eea
At this order we have two extra terms with coefficients $c^{(3)}_1$ and
$c^{(3)}_2$, which are arbitrary at this order:
\beq
\wh{E}_{\mu\nu}|_{(E_{**})^3}=e_{\mu\nu}|_{(E_{**})^3}
+c^{(3)}_1E_{\mu\nu}E_{\lambda\rho}E^{\lambda\rho}
+c^{(3)}_2E_{\mu\lambda}E_{\rho\nu}E^{\rho\lambda}.
\eeq
This satisfies (\ref{bscheij})-(\ref{bscheab}) at this order. Then,
\bea
\frac{1}{4}{\rm ln}\;{\rm det}(\delta_{ij}+\wh{E}_{ij}
 )\Bigg|_{(E_{**})^3} & = &
\frac{1}{4}\left(\wh{E}_{ii}-\Half\wh{E}_{ij}\wh{E}_{ji}
 +\frac{1}{3}\wh{E}_{ij}\wh{E}_{jk}\wh{E}_{ki}
 \right)\Bigg|_{(E_{**})^3} \nn
& = & \frac{1}{4}\Bigg(
 \frac{1}{4}E_i^{\;\;\lambda}E_\lambda^{\;\;\rho}E_{\rho i}
 -\frac{1}{4}E_i^{\;\;\lambda}E_{\lambda j}E_{ji}
 -\frac{1}{4}E_{ij}E_j^{\;\;\lambda}E_{\lambda i} \nn
& & +\frac{1}{3}E_{ij}E_{jk}E_{ki}
 +c^{(3)}_1E_{ii}E_{\lambda\rho}E^{\lambda\rho}
 +c^{(3)}_2E_{i\lambda}E_{\rho i}E^{\rho\lambda}\Bigg) \nn
& = & \frac{1}{48}E_\mu^{\;\;\nu}E_\nu^{\;\;\lambda}E_\lambda^{\;\;\mu}
-\frac{1}{16}E_i^{\;\; j}E_j^{\;\; a}E_a^{\;\; i}
-\frac{1}{16}E_a^{\;\; b}E_b^{\;\; c}E_c^{\;\; a} \nn
& & +\frac{1}{4}c^{(3)}_1
 (E_\lambda^{\;\;\lambda}-E_a^{\;\; a})E_{\mu\nu}E^{\mu\nu} \nn
& & +\frac{1}{4}c^{(3)}_2(E_\lambda^{\;\;\mu}E^{\nu\lambda}E_{\nu\mu}
 -E_a^{\;\;\mu}E^{\nu a}E_{\nu\mu}).
\eea
Therefore, by adding T-duality invariant terms, we obtain the following
covariant expression of $\wh{\phi}$:
\bea
\wh{\phi}|_{(E_{**})^3} & = &
\frac{1}{48}E_\mu^{\;\;\nu}E_\nu^{\;\;\lambda}E_\lambda^{\;\;\mu}
+\frac{1}{4}c^{(3)}_1E_\lambda^{\;\;\lambda}E_{\mu\nu}E^{\mu\nu}
+\frac{1}{4}c^{(3)}_2E_\lambda^{\;\;\mu}E^{\nu\lambda}E_{\nu\mu} \nn
& = & -\Half{\rm ln}\;{\rm det}
 \left(\delta_\mu^{\;\;\nu}-\Half E_\mu^{\;\;\nu}\right)\Bigg|_{(E_{**})^3}
+\frac{1}{4}c^{(3)}_1E_\lambda^{\;\;\lambda}E_{\mu\nu}E^{\mu\nu}
+\frac{1}{4}c^{(3)}_2E_\lambda^{\;\;\mu}E^{\nu\lambda}E_{\nu\mu}.
\eea
Order $(E_{**})^4$ contribution is calculated similarly:
\bea
\wh{E}_{\mu\nu}|_{(E_{**})^4} & = & e_{\mu\nu}|_{(E_{**})^4} \nn
& & +c^{(3)}_1E_\mu^{\;\;\lambda}E_{\lambda\nu}E_{\rho\sigma}E^{\rho\sigma}
 +\Half c^{(3)}_2(E_{\mu\lambda}E_{\rho\nu}E^{\lambda\sigma}E^\rho_{\;\;\sigma}
 +E_{\mu\lambda}E_{\rho\nu}E^{\sigma\lambda}E_\sigma^{\;\;\rho}), \\
\wh{\phi}|_{(E_{**})^4} & = & 
 -\Half{\rm ln}\;{\rm det}
 \left(\delta_\mu^{\;\;\nu}-\Half E_\mu^{\;\;\nu}\right)\Bigg|_{(E_{**})^4}
\nn & &
 +\Half c^{(3)}_1E_{\mu\nu}E^{\nu\mu}E_{\lambda\rho}E^{\lambda\rho}
 +\Half c^{(3)}_2E_{\mu\rho}E_\nu^{\;\;\mu}E^{\rho\lambda}E^\nu_{\;\;\lambda}
\nn & &
 +c^{(4)}_1E_{\mu\nu}E^{\mu\nu}E_{\lambda\rho}E^{\lambda\rho}
 +c^{(4)}_2E_{\mu\rho}E_\nu^{\;\;\rho}E^{\mu\lambda}E^\nu_{\;\;\lambda},
\eea
where $c^{(4)}_1$ and $c^{(4)}_2$ are arbitrary constants.
Note that some coefficients of order $(E_{**})^4$ terms are determined by
lower order coefficients $c^{(3)}_1$ and $c^{(3)}_2$.

In this manner we can continue this order by order analysis.
In general, new coefficients appear at order $(E_{**})^{2n+1}$
for $\wh{E}_{\mu\nu}$, and at order $(E_{**})^{2n}$ for $\wh{\phi}$.
Since new terms in $\wh{\phi}$ do not affect higher order computation,
their coefficients are left undetermined. Therefore $\wh{\phi}$ has
infinitely many undetermined coefficients. Although the above low order
computation does not prove that $c^{(3)}_1$ and $c^{(3)}_2$ are completely
arbitrary after taking full order effect into account, it seems that
$\wh{E}_{\mu\nu}$ also has infinitely many undetermined coefficients.
Since it seems difficult to find general form of this kind of term,
we do not investigate them further.

We have shown that some coefficients in the field redefinitions can be
determined analytically, just by assuming the correspondence of the 
T-duality transformations of the two sides and the covariance of the terms
in the field redefinitions. Therefore those terms are universal and
irrelevant of the detail of the definition of interaction terms and the
integrating-out procedure.

%%%%%%%%%%%%%%%%%%%%%%%%%%%%%%%%%%%%%%%%%%%%%%%%%%%%%%%%%%%%%%%%%%%
%%%%%%%%%%%%%%%%%%%%%%%%%%%%%%%%%%%%%%%%%%%%%%%%%%%%%%%%%%%%%%%%%%%
\section{Solutions in Closed String Field Theories I}

In this section we investigate solutions in
closed string field theories based on CFT for the flat spacetime,
by employing the method of section 2 and \cite{mchsht}. Then we show that
by the field redefinitions these solutions are identified with an $\ap$-exact
solution in the effective theory known as (generalized) chiral null model.

We separate spacetime coordinates $x^\mu$ into $x^\pm$ and $x^i$.
Let us consider the following configuration in the bosonic string case:
\bea
E_{+\mu}=0,\quad E_{i\mu}=0,\quad E_{-\mu}=E_{-\mu}(k_-,k_i), \nn
E^{(1)}=0,\quad E^{(2)}=0,\quad E^{(3)}_\mu=0,\quad E^{(4)}_\mu=0.
\eea
In terms of string field,
\beq
\Phi_0=\int\frac{d^{26}k}{(2\pi)^{26}}\frac{1}{\ap}
 \Big[E_{-+}(k_-,k_i)V^{-+}+E_{--}(k_-,k_i)V^{--}
 +E_{-i}(k_-,k_i)V^{-i}\Big],
\eeq
where $V^{\mu\nu}=c\bar{c}\p X^\mu\bar{\p}X^\nu e^{i(k_-X^-+k_iX^i)}$.
Note that the structure of the left mover of this configuration is the same
as that of the solution in section 2.

Let us solve the linearized equation of motion $Q\Phi_0=0$.
From the equations of motion of $E^{(3)}_\mu$ and $E^{(4)}_\mu$,
which correspond to Siegel gauge condition,
\beq
k^\nu E_{\nu\mu}=k^\nu E_{\mu\nu}=0\Rightarrow -k_-E_{-+}+k_iE_{-i}=0.
\label{sftgfix1}
\eeq
From the equations of motion of $E_{\mu\nu}$,
\beq
k_ik^iE_{\mu\nu}=0.
\label{sfteom1}
\eeq
Then equations of motion of $E^{(1)}$ and $E^{(2)}$ are trivially satisfied.

Full order solution can be obtained by expanding $\Phi$ in some parameter $g$:
$\Phi=g\Phi_0+g^2\Phi_1+g^3\Phi_2+\dots$.
Then the equation of motion
\beq
0=Q\Phi+\sum_{n=2}^\infty\frac{1}{n!}[\Phi^n]
\label{bsfteom1}
\eeq
is decomposed into contributions from each order in $g$:
\beq
\Delta_N=Q\Phi_N+\sum_{n=2}^{N+1}\frac{1}{n!}\sum_{\stackrel{
 \mbox{$\scriptstyle 0\leq N_1, N_2,\dots, N_n\leq N-1$}}
{\mbox{$\scriptstyle N_1+N_2+\dots+N_n=N-n+1$}}}
[\Phi_{N_1}, \Phi_{N_2}, \dots, \Phi_{N_n}],
\label{bsfteom2}
\eeq
where we introduced source term $\Delta=g\Delta_0+g^2\Delta_1+\dots$ in
the left hand side of (\ref{bsfteom1}). For the definition of the closed
string field product $[\cdot,\cdot,\dots,\cdot]$, see e.g. \cite{z1}.
Coordinate expressions of components in $\Delta_0$ are delta functions
so that $Q\Phi_0=\Delta_0$ gives correct linearized equations with delta
function source terms.
$\Delta$ should also satisfy $(L_0-\bar{L}_0)\Delta=(b_0-\bar{b}_0)\Delta=0$.
We demand $\Phi_N$ and $\Delta_N$ for $N\geq 1$ satisfy these conditions and 
$(b_0+\bar{b}_0)\Phi_N=(b_0+\bar{b}_0)\Delta_N=0$. i.e.
\bea
 & & b_0\Phi_N=\bar{b}_0\Phi_N=0, \\
 & & b_0\Delta_N=\bar{b}_0\Delta_N=0.
\eea
This condition on $\Delta_N$ means that physical massive modes have no source.
Then the equations (\ref{bsfteom2}) are solved order by order:
\bea
\Phi_N & = & -\frac{b_0+\bar{b}_0}{L_0+\bar{L}_0}\sum_{n=2}^{N+1}
 \frac{1}{n!}\sum_{\stackrel{
 \mbox{$\scriptstyle 0\leq N_1, N_2,\dots, N_n\leq N-1$}}
{\mbox{$\scriptstyle N_1+N_2+\dots+N_n=N-n+1$}}}
[\Phi_{N_1}, \Phi_{N_2}, \dots, \Phi_{N_n}],
\label{bsftsol} \\
\Delta_N & = & -\frac{b_0+\bar{b}_0}{L_0+\bar{L}_0}\sum_{n=2}^{N+1}
 \frac{1}{(n-1)!}\sum_{\stackrel{
 \mbox{$\scriptstyle 0\leq N_1, N_2,\dots, N_n\leq N-1$}}
{\mbox{$\scriptstyle N_1+N_2+\dots+N_n=N-n+1$}}}
[\Delta_{N_1}, \Phi_{N_2}, \dots, \Phi_{N_n}],
\label{bsftsrc}
\eea
i.e. $\Phi_N$ is expressed by lower order $\Phi_M$, and $\Delta_N$
is expressed by lower order $\Phi_M$ and $\Delta_M$. From (\ref{bsftsrc}),
$\Delta_N=0$ for any $N$ if $\Delta_0=0$.
(\ref{bsftsol}) can be shown by acting $b_0+\bar{b}_0$ on (\ref{bsfteom2})
and using $\{Q, b_0+\bar{b}_0\}=L_0+\bar{L}_0$.
(\ref{bsftsrc}) can be shown by plugging (\ref{bsftsol}) into
(\ref{bsfteom2}):
\bea
\Delta_N & = & \frac{b_0+\bar{b}_0}{L_0+\bar{L}_0}Q\sum_{n=2}^{N+1}
 \frac{1}{n!}\sum_{\stackrel{
 \mbox{$\scriptstyle 0\leq N_1, N_2,\dots, N_n\leq N-1$}}
 {\mbox{$\scriptstyle N_1+N_2+\dots+N_n=N-n+1$}}}
 [\Phi_{N_1}, \Phi_{N_2}, \dots, \Phi_{N_n}] \nn
 & = & -\frac{b_0+\bar{b}_0}{L_0+\bar{L}_0}\Bigg(
 \sum_{n=2}^{N+1}\sum_{\stackrel{
 \mbox{$\scriptstyle 0\leq N_1, N_2,\dots, N_n\leq N-1$}}
 {\mbox{$\scriptstyle N_1+N_2+\dots+N_n=N-n+1$}}}
 \frac{1}{(n-1)!}[Q\Phi_{N_1}, \Phi_{N_2}, \dots, \Phi_{N_n}] \nn
& & +\sum_{n=3}^{N+1}\sum_{\stackrel{
 \mbox{$\scriptstyle 0\leq N_1, N_2,\dots, N_n\leq N-1$}}
 {\mbox{$\scriptstyle N_1+N_2+\dots+N_n=N-n+1$}}}
 \sum_{m=1}^{n-2}\frac{1}{m!(n-m)!}[\Phi_{N_1}, \dots, \Phi_{N_m},
 [\Phi_{N_{m+1}}, \dots, \Phi_{N_n}]]\Bigg),
\eea
where we used the following identity (see e.g. \cite{z1}):
\bea
Q[\Phi_{N_1}, \Phi_{N_2}, \dots, \Phi_{N_n}] & = & 
-[Q\Phi_{N_1}, \Phi_{N_2}, \dots, \Phi_{N_n}]
-[\Phi_{N_1}, Q\Phi_{N_2}, \dots, \Phi_{N_n}]-\dots \nn
& & -[\Phi_{N_1}, \Phi_{N_2}, \dots, Q\Phi_{N_n}] \nn
& & -\sum_{\{i_l,j_k\}}[\Phi_{N_{i_1}}, \dots, \Phi_{N_{i_l}},
 [\Phi_{N_{j_1}}, \dots, \Phi_{N_{j_k}}]].
\eea
The sum $\sum_{\{i_l,j_k\}}$ runs over all different splittings of the
set $\{1,2,\dots, n\}$ into a first group $\{{i_1},\dots, {i_l}\}$ $(l\geq 1)$
and a second group $\{{j_1},\dots, {j_k}\}$ $(k\geq 2)$, regardless of the
order of the integers.

Then by eliminating $Q\Phi_{N_i}$ using (\ref{bsfteom2}),
\bea
\Delta_N & = & -\frac{b_0+\bar{b}_0}{L_0+\bar{L}_0}\Bigg(
 \sum_{n=2}^{N+1}\sum_{\stackrel{
 \mbox{$\scriptstyle 0\leq N_1, N_2,\dots, N_n\leq N-1$}}
 {\mbox{$\scriptstyle N_1+N_2+\dots+N_n=N-n+1$}}}
 \frac{1}{(n-1)!}[\Delta_{N_1}, \Phi_{N_2}, \dots, \Phi_{N_n}] \nn
& & -\sum_{n=2}^{N+1}\sum_{\stackrel{
 \mbox{$\scriptstyle 0\leq N_1, N_2,\dots, N_{n-1}\leq N-1,
 1\leq N_n\leq N-1$}}
 {\mbox{$\scriptstyle N_1+N_2+\dots+N_n=N-n+1$}}} \nn
& & \times\sum_{m=2}^{N_n+1}
 \sum_{\stackrel{
 \mbox{$\scriptstyle 0\leq M_1, M_2,\dots, M_m\leq N_n-1$}}
 {\mbox{$\scriptstyle M_1+M_2+\dots+M_m=N_n-m+1$}}}
 \frac{1}{(n-1)!m!}[\Phi_{N_1}, \dots, \Phi_{N_{n-1}},
 [\Phi_{M_1}, \dots, \Phi_{M_m}]] \nn
& & +\sum_{n=3}^{N+1}\sum_{\stackrel{
 \mbox{$\scriptstyle 0\leq N_1, N_2,\dots, N_n\leq N-1$}}
 {\mbox{$\scriptstyle N_1+N_2+\dots+N_n=N-n+1$}}}
 \sum_{m=1}^{n-2}\frac{1}{m!(n-m)!}[\Phi_{N_1}, \dots, \Phi_{N_m},
 [\Phi_{N_{m+1}}, \dots, \Phi_{N_n}]]\Bigg).
\eea
By a careful rearrangement of the summation we can show that the second term
in the above equation is equal to minus the third term,
and (\ref{bsftsrc}) follows.

Since $g$ can be absorbed into $E_{\mu\nu}$ by the rescaling
$\Phi_0\rightarrow\Phi_0/g$, henceforth we put $g=1$.

Note that Fock space representation of the left moving part of $\Phi_0$ has
only $\alpha^-_{-m}$, and has no $k_+$ dependence. Therefore the difference
$n_--n_+$ of the number of
$\alpha^-_{-m}$ and $\alpha^+_{-m}$ of the left moving part of $\Phi_N$
is greater than or equal to $N+1$. Therefore the minimum level of the left
moving part of $\Phi_N$ increases as $N$ increases. Then by the level
matching condition $(L_0-\bar{L}_0)\Phi=0$, the minimum level of the
right mover also increases.
This fact can be proven in a similar way as in \cite{mchsht}.

Therefore this solution has properties similar to those in section 2:
Tachyon component of $\Phi$ is exactly zero, massless components have no
higher correction, and each coefficient of massive Fock space state receives
corrections from finitely many $\Phi_N$. Inverses of $L_0+\bar{L}_0$ are
well-defined.

By the argument similar to that of \cite{mchsht},
we can see that $\Phi_1$ is well defined and smooth everywhere, even if
$\Phi_0$ has singularities due to the effect of the source terms.
As has been pointed out in \cite{kz}, $\Phi_N$ can be written in terms of
$(N+2)$-point off-shell amplitudes. Although it is technically difficult to
compute 4-point or higher amplitude in closed string field theory, we expect
that higher $\Phi_N$ are also well-defined and smooth everywhere. 

Next we consider analogous solution in the heterotic string field theory.
The linearized solution is
\bea
\Phi_0 & = & \int\frac{d^{10}k}{(2\pi)^{10}}
\frac{i}{\sqrt{2\ap}}\Big[E_{-+}(k_-,k_i)V^{-+}+E_{--}(k_-,k_i)V^{--}
 +E_{-i}(k_-,k_i)V^{-i}\Big],
\eea
where $V^{\mu\nu}=\xi c\psi^\mu e^{-\phi}\bar{c}\bar{\p}X^\nu
e^{i(k_-X^-+k_iX^i)}$.
The linearized equation of motion $\eta_0Q\Phi_0=0$ reduces to the same
equations (\ref{sftgfix1}) and (\ref{sfteom1}) as in the bosonic case.

The fully nonlinear equation of motion of this theory is
\bea
0 & = & \eta_0\bar{\Psi}_Q \nn
 & = & \eta_0\left(Q\Phi+\Half[\Phi, Q\Phi]+\frac{1}{3!}[\Phi, Q\Phi, Q\Phi]
 +\frac{1}{3!}[\Phi, [\Phi, Q\Phi]]+O(\Phi^4)\right) \nn
 & = & \eta_0Q\Phi+\eta_0Z.
\label{hsfteom1}
\eea
For the definition of $\bar{\Psi}_Q$, see \cite{boz}.\footnote{
$V(1)$ in \cite{boz} is equal to $\kappa^{-1}\Phi$.}
$Z$ is defined by $\bar{\Psi}_Q=Q\Phi+Z$, and is quadratic or higher in $\Phi$.
As in the bosonic case, $\Phi$ is expanded in a parameter $g$:
$\Phi=g\Phi_0+g^2\Phi_1+\dots$. Accordingly $Z$ is also expanded:
$Z=g^2Z_1+g^3Z_2+\dots (Z_0=0)$.
$Z_N$ consists of $\Phi_M$ with $M=0,1,\dots,N-1$.
Then (\ref{hsfteom1}) is decomposed into
\beq
\Delta_N=\eta_0Q\Phi_N+\eta_0Z_N, \label{hsfteom}
\eeq
where we introduced source term $\Delta=g\Delta_0+g^2\Delta_1+\dots$ in the
left hand side of (\ref{hsfteom1}).
We demand that $\Phi_N$ and $\Delta_N$ satisfy
\bea
& & b_0\Phi_N=\bar{b}_0\Phi_N=\wt{G}^-_0\Phi_N=0, \label{hsftgfix} \\
& & b_0\Delta_N=\bar{b}_0\Delta_N=0.
\eea
(\ref{hsftgfix}) means that when we consider integrating-out procedure
described in section 3 we take this partial gauge fixing condition
for massive modes.

(\ref{hsfteom}) is solved order by order by the following:
\bea
\Phi_N & = & -\frac{\wt{G}^-_0}{L_0}
\frac{b_0+\bar{b}_0}{L_0+\bar{L}_0}\eta_0Z_N,
\label{hsftsol} \\
\Delta_N & = & -\frac{b_0+\bar{b}_0}{L_0+\bar{L}_0}
 \sum_{n=1}^N\frac{1}{n!}\sum_{\stackrel{
 \mbox{$\scriptstyle 0\leq N_1, N_2,\dots, N_{n+1}\leq N-1$}}
 {\mbox{$\scriptstyle N_1+N_2+\dots+N_{n+1}=N-n$}}} \nn
 & & \times[\Delta_{N_1},
Q\Phi_{N_2}+Z_{N_2}, Q\Phi_{N_3}+Z_{N_3},\dots, Q\Phi_{N_{n+1}}+Z_{N_{n+1}}].
\label{hsftsrc}
\eea
(\ref{hsftsol}) can be derived by acting $(b_0+\bar{b}_0)\wt{G}^-_0$ on
(\ref{hsfteom}). (\ref{hsftsrc}) can be derived as follows.
Plugging (\ref{hsftsol}) into (\ref{hsfteom}), we obtain
\beq
\Delta_N=\frac{b_0+\bar{b}_0}{L_0+\bar{L}_0}Q\eta_0Z_N. \label{hsftsrc2}
\eeq
On the other hand, from the following identity \cite{boz}:
\beq
Q\bar{\Psi}_Q+\sum_{n=2}^\infty\frac{1}{n!}[\bar{\Psi}_Q^n]=0,
\eeq
we obtain
\beq
Q\eta_0\bar{\Psi}_Q=-\sum_{n=1}^\infty\frac{1}{n!}[\eta_0\bar{\Psi}_Q,
\bar{\Psi}_Q^n].
\eeq
Extracting order $g^{N+1}$ contribution of this identity and 
using (\ref{hsfteom}) in the right hand side,
\beq
Q\eta_0Z_N=-\sum_{n=1}^N\frac{1}{n!}\sum_{\stackrel{
 \mbox{$\scriptstyle 0\leq N_1, N_2,\dots, N_{n+1}\leq N-1$}}
 {\mbox{$\scriptstyle N_1+N_2+\dots+N_{n+1}=N-n$}}}
[\Delta_{N_1},
Q\Phi_{N_2}+Z_{N_2}, Q\Phi_{N_3}+Z_{N_3},\dots, Q\Phi_{N_{n+1}}+Z_{N_{n+1}}].
\eeq
From this and (\ref{hsftsrc2}), (\ref{hsftsrc}) follows.

This solution has a similar property to the bosonic one:
Let $n_\pm$ be the sum of the numbers of $\alpha^\pm_{-m}$ and $\psi^\pm_{-r}$
in the Fock space representation of the left moving part of $\Phi_N$,
then $n_--n_+\geq N+1$.

Therefore massless components have no higher correction, 
each coefficient of massive Fock space state receives
corrections from finite number of $\Phi_N$, and
inverses of $L_0+\bar{L}_0$ are well-defined.

Is there a solution of the effective theory corresponding to these
string field theory solutions? Here is a candidate known as (generalized)
chiral null model\cite{ht,cmp}.
This model gives solutions of both bosonic and heterotic effective
theory, and consists of nontrivial string metric
$ds^2=\wh{g}_{\mu\nu}dx^\mu dx^\nu$, B-field $\wh{B}_{\mu\nu}$
and dilaton $\wh{\phi}$:
\bea
ds^2 & = & \wh{F}dx^-dx^++\wh{F}\wh{K}(dx^-)^2+2\wh{F}\wh{A}_idx^-dx^i
 +dx^idx^i, \\
\wh{B}_{-+} & = & \Half\wh{F}+1, \\
\wh{B}_{-i} & = & \wh{F}\wh{A}_i, \\
\wh{\phi} & = & \wh{\phi}_0(x^-)+\Half{\rm ln}(-\wh{F}),
\label{effphi1}
\eea
where $\wh{F}$, $\wh{K}$, $\wh{A}_i$ are functions of $x^-$ and $x^i$:
$\wh{F}=\wh{F}(x^-,x^i)$, $\wh{K}=\wh{K}(x^-,x^i)$,
$\wh{A}_i=\wh{A}_i(x^-,x^i)$, and $\wh{\phi}_0$ is a function of $x^-$
\footnote{We can further introduce linear dilaton term
which shifts the central charge\cite{ht}.}.
In matrix notation,
\beq
\wh{E}_{\mu\nu}\equiv\wh{h}_{\mu\nu}+\wh{B}_{\mu\nu}
=\bordermatrix{ & - & + & i \cr
- & \wh{F}\wh{K} & \wh{F}+2 & 2\wh{F}\wh{A}_i \cr
+ & 0 & 0 & 0 \cr
i & 0 & 0 & 0 \cr}.
\label{effsol1}
\eeq
By coordinate transformation $x^+\rightarrow x^+-2\eta(x^-,x^i)$,
$\wh{K}$ and $\wh{A}_i$ are ``gauge transformed'':
\beq
\wh{K}\rightarrow\wh{K}+2\p_-\eta,\quad\wh{A}_i\rightarrow\wh{A}_i+\p_i\eta.
\eeq
Equations of motion of two derivative truncation of the effective theory
are reduced to the following equations, which determine
$\wh{F}$, $\wh{K}$ and $\wh{A}_i$:
\bea
0 & = & \p_i\p^i\wh{F}^{-1},
\label{effeom1} \\
0 & = & \p^j\wh{{\mathcal F}}_{ij}
 +\wh{F}^{-1}e^{2\wh{\phi}}\p_-(\wh{F}^{-1}e^{-2\wh{\phi}}\p_i\wh{F}),
\label{effeom2} \\
0 & = & -\Half\p_i\p^i\wh{K}+\p_-\p^i\wh{A}_i
 +2\wh{F}^{-1}\p_-^2\wh{\phi}_0
 +\wh{F}^{-1}e^{2\wh{\phi}}\p_-(\wh{F}^{-1}e^{-2\wh{\phi}}\p_-\wh{F}),
\label{effeom3}
\eea
where $\wh{{\mathcal F}}_{ij}=\p_i\wh{A}_j-\p_j\wh{A}_i$.

Important examples of this class of solution are configurations of
infinitely extended macroscopic F-strings\cite{dghr},
and F-strings with waves on them\cite{cmp,dghw}.
To obtain these solutions we have to introduce delta function source terms
in the left hand sides of (\ref{effeom1}), (\ref{effeom2}) and
(\ref{effeom3}).

Although we introduced this configuration as a solution of two derivative
truncation of the effective theory, in fact it has been shown that
this is an $\ap$-exact solution\cite{ht,cmp}.

Equations (\ref{sftgfix1}) and (\ref{sfteom1}) are linear. Therefore we
have to linearize the equations (\ref{effeom1}), (\ref{effeom2}) and
(\ref{effeom3}) in order to see correspondence between
our solutions. We assume that $\wh{\phi}_0$ is a constant, and
$\wh{A}_i$ satisfy the gauge fixing condition
\beq
-\p_-\wh{F}^{-1}+\p^i\wh{A}_i=0,
\label{effgfix1}
\eeq
which is analogous to (\ref{sftgfix1}). Then (\ref{effeom1}), (\ref{effeom2}),
(\ref{effeom3}) and (\ref{effphi1}) are
\bea
0 & = & \p_i\p^i\wh{F}^{-1} \label{effeom4}, \\
0 & = & \p_i\p^i\wh{A_j} \label{effeom5}, \\
0 & = & \p_i\p^i\wh{K} \label{effeom6}, \\
\wh{\phi} & = & \wh{\phi}_0+\Half{\rm ln}(-\wh{F}). \label{effphi2} 
\eea
We can see the similarity between two solutions: $E_{\mu\nu}$ and
$\wh{E}_{\mu\nu}$ has the same nonzero entries and the same coordinate
dependence, and satisfy similar linear equations and gauge fixing conditions.
To make the identification clearer,
we define $F(x^-,x^i)$, $K(x^-,x^i)$ and $A_i(x^-,x^i)$ as follows.
\beq
E_{\mu\nu}\equiv\left(\begin{array}{ccc}
-2K & F+2 & -4A_i \\
0 & 0 & 0 \\ 0 & 0 & 0
\end{array}\right).
\label{sftsol1}
\eeq
Then the gauge fixing condition (\ref{sftgfix1}) is
\beq
0=\frac{1}{4}\p_-F+\p^iA_i, \label{sftgfix2}
\eeq
Equations of motion and $\phi$ are
\bea
0 & = & \p_i\p^iF, \label{sfteom2} \\
0 & = & \p_i\p^iA_j, \\
0 & = & \p_i\p^iK, \\
\phi & = & -\frac{1}{4}(F+2). \label{sftphi1}
\eea
Although these are very similar to (\ref{effgfix1}), (\ref{effeom4}),
(\ref{effeom5}), (\ref{effeom6}) and (\ref{effphi2}), we notice some
difference. This is because for these solutions $E_{\mu\nu}$ and $\phi$
are not equal to $\wh{E}_{\mu\nu}$ and $\wh{\phi}$ respectively, unlike
the open string case in section 2. Some correction terms in the field
redefinitions remain nonzero.

Note that the field redefinitions are directly available because we know that
$E_{\mu\nu}$ and $\phi$ on the string field theory side and
$\wh{E}_{\mu\nu}$ and $\wh{\phi}$ on the effective theory side have no
higher order correction.
But before applying the field redefinitions,
let us determine the relation between $F$, $K$, $A_i$ and $\wh{F}$,
$\wh{K}$, $\wh{A}_i$. Roughly speaking $\wh{F}$ is inverse of $F$ as we can
see from (\ref{effgfix1}), (\ref{effeom4}), (\ref{sftgfix2}) and
(\ref{sfteom2}). To determine the precise
relation, first note that $F+2$ and $\wh{F}+2$ are the deviations from the
flat metric, and at the linearized order in $F+2$ our solutions should be
the same. We assume that $\wh{F}^{-1}=a+bF$. Then,
\bea
\wh{F}+2 & = & 2+(a-2b+b(F+2))^{-1} \nn
 & = & 2+(a-2b)^{-1}-b(a-2b)^{-2}(F+2)+O((F+2)^2).
\eea
Therefore,
\beq
2+(a-2b)^{-1}=0,\quad -b(a-2b)^{-2}=1,
\eeq
which are solved by $a=-1$ and $b=-1/4$. Hence
\beq
\wh{F}=-\frac{1}{1+\frac{1}{4}F},
\eeq
and
\beq
\wh{K}=K,\quad\wh{A}_i=A_i.
\eeq
Then we can explicitly see that the equations of motion on both sides are
equivalent, and the gauge fixing conditions are the same.

Now we are ready to apply the field redefinitions and see whether the solutions
are really the same. Let us start with considering derivative terms.
Note that when we claim that a solution of two-derivative truncation of the
fully corrected effective theory is an $\ap$-exact solution,
we choose some particular form of higher derivative terms. 
This means that correspondingly the ambiguity which we discussed in section 3
is fixed in some particular way.

Since $F$, $K$ and $A_i$ are restricted only by the Laplace equations
and the gauge fixing condition, they and their derivatives
$\p_* F$, $\p_*\p_* F$, $\dots$, 
$\p_* K$, $\p_*\p_* K$, $\dots$, 
$\p_* A_i$, $\p_*\p_* A_i$, $\dots$ are independent functions.
Although $F$ and $A_i$ are related by (\ref{sftgfix2}),
they still have enough degrees of freedom which enable us
to regard them and their derivatives as independent
as we can see by taking $F$ independent of $x^-$.
Then, terms with derivatives in the field redefinitions should cancel each
other when we plug our string field theory solution into it, because
$\wh{E}_{\mu\nu}$ and $\wh{\phi}$ on the effective theory side does not
contain derivatives of $F$, $K$ and $A_i$.

Assuming this, let us consider the remaining terms i.e. those without
derivatives. Since $T=0$ in the bosonic case and $A^a_\mu=0$ in the heterotic
case, we do not have to take terms with $T$ and $A^a_\mu$ into account.
More importantly, for our string field theory solution 1-1 contraction
of two $E_{**}$ vanishes:
$E_{\lambda *}E_{\;\; *}^\lambda=0$, which means that terms we
could not determine in (\ref{eresult}) and (\ref{phiresult}) are zero.
Then the remaining terms are
\bea
\wh{E}_{\mu\nu} & = & e_{\mu\nu}, \\
\wh{\phi} & = & c+\phi-\frac{1}{4}E_\mu^{\;\;\mu}
 -\Half{\rm ln}\;{\rm det}\left(\delta_\mu^{\;\;\nu}
 -\Half E_\mu^{\;\;\nu}\right).
\eea
We can see that plugging (\ref{sftsol1}) and
(\ref{sftphi1}) into the above, (\ref{effsol1}) and (\ref{effphi2})
are reproduced exactly, with the following identification between the
constants:
\beq
\wh{\phi}_0=c-\Half {\rm ln} 2.
\eeq

%%%%%%%%%%%%%%%%%%%%%%%%%%%%%%%%%%%%%%%%%%%%%%%%%%%%%%%%%%%%%%%%%%%
%%%%%%%%%%%%%%%%%%%%%%%%%%%%%%%%%%%%%%%%%%%%%%%%%%%%%%%%%%%%%%%%%%%
\section{Solutions in Closed String Field Theories II}

In this section we consider another type of string field theory solution,
and identify it with another solution in the effective theory: pp-wave
solution with nontrivial B-field, which is known to be $\ap$-exact under
some condition.

We split $x^\mu$ into $x^\pm$, $x^{i_1}$ and $x^{i_2}$.
Index $i$ runs over both ranges of $i_1$ and $i_2$.
We give linearized solutions different from the previous section.
In the bosonic case,
\bea
\Phi_0 & = & \int\frac{d^{26}k}{(2\pi)^{26}}\frac{1}{\ap}
 \Big[E_{--}(k_-,k_{i_1},k_{i_2})V^{--}(k_-,k_{i_1},k_{i_2}) \nn
 & & +E_{-i_1}(k_-,k_{i_2})V^{-i_1}(k_-,0,k_{i_2})
 +E_{i_1-}(k_-,k_{i_2})V^{i_1-}(k_-,0,k_{i_2})\Big],
\eea
where $V^{\mu\nu}(k_-,k_{i_1},k_{i_2})
=c\bar{c}\p X^\mu\bar{\p}X^\nu e^{i(k_-X^-+k_{i_1}X^{i_1}+k_{i_2}X^{i_2})}$.

In the heterotic case,
\bea
\Phi_0 & = & \int\frac{d^{10}k}{(2\pi)^{10}}\frac{i}{\sqrt{2\ap}}
 \Big[E_{--}(k_-,k_{i_1},k_{i_2})V^{--}(k_-,k_{i_1},k_{i_2}) \nn
 & & +E_{-i_1}(k_-,k_{i_2})V^{-i_1}(k_-,0,k_{i_2})
 +E_{i_1-}(k_-,k_{i_2})V^{i_1-}(k_-,0,k_{i_2})\Big],
\eea
where $V^{\mu\nu}(k_-,k_{i_1},k_{i_2})
=\xi c\psi^\mu e^{-\phi}\bar{c}\bar{\p}X^\nu
e^{i(k_-X^-+k_{i_1}X^{i_1}+k_{i_2}X^{i_2})}$.

In both cases the linearized equations reduce to
\beq
k_ik^iE_{\mu\nu}=0,
\eeq
and the solutions for fully nonlinear equations of motion are constructed
in the same way as in the previous section. Roughly speaking, since $\Phi_0$
has no $X^+$ and $\psi^+$, numbers of $X^-$ or $\psi^-$ in either left or
right mover always increase when we take string field product of
$V^{--}(k_-,k_{i_1},k_{i_2})$ and $\Phi_0$, which means the minimum levels of
both left and right movers increase by the level matching condition.
When we take string field product of two $V^{-{i_1}}(k_-,0,k_{i_2})$, or
two $V^{{i_1}-}(k_-,0,k_{i_2})$, numbers of $X^-$ or $\psi^-$ in either
left mover or right mover increase, which again means the minimum levels of
both left and right mover increase.
When we take string field product of $V^{-{i_1}}(k_-,0,k_{i_2})$ and
$V^{{i_1}-}(k_-,0,k_{i_2})$, numbers of $X^-$, $\psi^-$, $X^{i_1}$ or
$\psi^{i_1}$ in both left mover and right mover increase.
Therefore the minimum level of product of $\Phi_0$ increases as we multiply
more and more $\Phi_0$.

The precise statement is the following:
Let $n_\pm$ be the sum of the numbers of $\alpha^\pm_{-m}$ and $\psi^\pm_{-r}$
in the Fock space representation of the left moving part of $\Phi_N$,
$n_1$ be the sum of the numbers of $\alpha^{i_1}_{-m}$ and $\psi^{i_1}_{-r}$
in the left moving part of $\Phi_N$,
and let $\bar{n}_\pm$ and $\bar{n}_1$ be analogous numbers in the right
moving part. Then $(n_-+\bar{n}_-)-(n_++\bar{n}_+)\geq N+1$.
In addition, for $\Phi_1$, $n_-+n_1-n_+\geq 2$ or
$\bar{n}_-+\bar{n}_1-\bar{n}_+\geq 2$.
Again these facts can be proven in a way similar to that in \cite{mchsht}.

For $N=1$, $n_-+n_1$ or $\bar{n}_-+\bar{n}_1$ is larger than 2, which means
$\Phi_1$ has no tachyon and massless components.
For $N\geq 2$, $2{\rm max}(n_-,\bar{n}_-)\geq n_-+\bar{n}_-
\geq N+1+n_++\bar{n}_+\geq N+1\geq 3$.
Therefore ${\rm max}(n_-,\bar{n}_-)\geq 2$, which again means that 
$\Phi_N$ has no tachyon and massless components.
Therefore this solution has the same properties as those of the solution in
the previous section.

Again we have a candidate for the solution of the effective theory
corresponding to this string field theory solution. It is the pp-wave
solution with nontrivial B-field:
\bea
ds^2 & = & -2dx^-dx^+-2\wh{K}(dx^-)^2-4\wh{A}_idx^-dx^i+dx^idx^i, \\
\wh{B}_{-i} & = & 2\wh{B}_i, \\
\wh{\phi} & = & \wh{\phi}_0.
\eea
where $\wh{\phi}_0$ is a constant, $\wh{K}=\wh{K}(x^-,x^i)$,
$\wh{A}_i=\wh{A}_i(x^-,x^i)$ and
$\wh{B}_i=\wh{B}_i(x^-,x^i)$.

Coordinate transformation $x^+\rightarrow x^+-2\eta(x^-,x^i)$ and gauge
transformation of B-field induce the following transformations.
\beq
\wh{K}\rightarrow\wh{K}+2\p_-\eta(x^-,x^i),\quad
\wh{A}_i\rightarrow\wh{A}_i+\p_i\eta(x^-,x^i),
\eeq
\beq
\wh{B}_i\rightarrow\wh{B}_i+\p_i\zeta(x^-,x^i).
\eeq
The equations of motion of two derivative truncation of the effective action
reduce to
\bea
0 & = & \p^j\wh{{\mathcal F}}_{ij}, \\
0 & = & \p^j\wh{{\mathcal G}}_{ij}, \\
0 & = & \p_i\p^i\wh{K}-2\p_-\p^i\wh{A}_i
 +\wh{{\mathcal F}}_{ij}\wh{{\mathcal F}}^{ij}
 -\wh{{\mathcal G}}_{ij}\wh{{\mathcal G}}^{ij},
\eea
where $\wh{{\mathcal F}}_{ij}=\p_i\wh{A}_j-\p_j\wh{A}_i$ and
$\wh{{\mathcal G}}_{ij}=\p_i\wh{B}_j-\p_j\wh{B}_i$.

In \cite{ht,tstl,dhh} it has been shown that this solution is $\ap$-exact
when $\wh{{\mathcal F}}_{ij}$ and $\wh{{\mathcal G}}_{ij}$ are independent of
$x^i$.

To match nonzero entries and coordinate dependence with our string field
theory solution, we put a restriction:
$\wh{A}_{i_1}=\wh{A}_{i_1}(x^-,x^{i_2})$, $\wh{A}_{i_2}=0$,
$\wh{B}_{i_1}=\wh{B}_{i_1}(x^-,x^{i_2})$ and $\wh{B}_{i_2}=0$.
Then the equations of motion are
\bea
0 & = & \p_{i_2}\p^{i_2}\wh{A}_{j_1}, \\
0 & = & \p_{i_2}\p^{i_2}\wh{B}_{j_1}, \\
0 & = & \p_i\p^i\left(\wh{K}+\wh{A}_{j_1}\wh{A}^{j_1}
 -\wh{B}_{j_1}\wh{B}^{j_1}\right).
\eea
In matrix notation,
\beq
\wh{E}_{\mu\nu}=\bordermatrix{ & - & + & i_1 & i_2 \cr
- & -2\wh{K} & 0 & -2\wh{A}_{i_1}+2\wh{B}_{i_1} & 0 \cr
+ & 0 & 0 & 0 & 0 \cr
i_1 & -2\wh{A}_{i_1}-2\wh{B}_{i_1} & 0 & 0 & 0 \cr
i_2 & 0 & 0 & 0 & 0 \cr}.
\label{effsolii}
\eeq
Accordingly, we define $K(x^-,x^i)$,
$A_{i_1}(x^-,x^{i_2})$ and $B_{i_1}(x^-,x^{i_2})$ as follows:
\beq
E_{\mu\nu}=\left(\begin{array}{cccc}
-2K & 0 & -2A_{i_1}+2B_{i_1} & 0 \\
 0 & 0 & 0 & 0 \\
 -2A_{i_1}-2B_{i_1} & 0 & 0 & 0 \\
 0 & 0 & 0 & 0 
\end{array}\right).
\label{sftsolii}
\eeq
Equations of motion and $\phi$ are
\bea
0 & = & \p_{i_2}\p^{i_2}A_{j_1}, \\
0 & = & \p_{i_2}\p^{i_2}B_{j_1}, \\
0 & = & \p_i\p^iK, \\
\phi & = & 0.
\eea

It is natural to identify these two solutions with the following relation:
\bea
\wh{K}+\wh{A}_{j_1}\wh{A}^{j_1}-\wh{B}_{j_1}\wh{B}^{j_1}
 & = & K, \\
\wh{A}_{j_1} & = & A_{j_1}, \\
\wh{B}_{j_1} & = & B_{j_1}.
\eea

Let us apply the field redefinitions and see if these solutions are really
the same. Since $K$, $\p_* K$, $\p_*\p_* K$, $\dots$, $A_{i_1}$,
$\p_* A_{i_1}$, $\p_*\p_* A_{i_1}$, $\dots$, $B_{i_1}$,
$\p_* B_{i_1}$, $\p_*\p_* B_{i_1}$, $\dots$ can be regarded as 
independent functions, we assume that 
terms with derivatives in the field redefinitions cancel each other.
We do not have to take terms with $T$ and $A^a_\mu$ into account
because $T=0$ in the bosonic case and $A^a_\mu=0$ in the heterotic case.
Unlike the solution in the previous section, 1-1 and 2-2 contractions do not
vanish. However, by straightforward calculation,
\bea
& & E_{\mu\nu}E^{\mu\nu}=0,\quad E_{\mu\nu}E^{\nu\mu}=0, \\ & &
E_{*\lambda}E^{\lambda\rho}E_{\rho *}=0,\quad
E_{*\lambda}E^{\rho\lambda}E_{\rho *}=0,\quad
E_{*\lambda}E^{\lambda\rho}E_{*\rho }=0,\quad
E_{\lambda *}E^{\lambda\rho}E_{\rho *}=0,
\eea
which means that terms cubic or higher in $E_{**}$, and terms with
coefficients left undetermined in section 4 are zero.
Then the remaining terms are
\bea
\wh{E}_{\mu\nu} & = & E_{\mu\nu}+\Half E_{\mu\lambda}E^\lambda_{\;\;\nu}, \\
\wh{\phi} & = & c+\phi.
\eea
By plugging (\ref{sftsolii}) into this (\ref{effsolii}) is reproduced, and
the relation between the constants is
\beq
\wh{\phi}_0=c.
\eeq
Since $\ap$-exactness of the solution on the effective theory side has
been proven only when $\wh{A}_{i_1}$ and $\wh{B}_{i_1}$ are linear in
$x^{i_2}$ (some more discussion has been given in \cite{rt}), and our
configuration is restricted by the weaker condition that
$\wh{A}_{i_1}$ and $\wh{B}_{i_1}$ satisfy Laplace equations,
it may not be possible to ignore higher derivative terms in the field
redefinitions in general.

%%%%%%%%%%%%%%%%%%%%%%%%%%%%%%%%%%%%%%%%%%%%%%%%%%%%%%%%%%%%%%%%%%%
%%%%%%%%%%%%%%%%%%%%%%%%%%%%%%%%%%%%%%%%%%%%%%%%%%%%%%%%%%%%%%%%%%%
\section{Discussion}

We have shown that some terms in the field redefinitions are determined
by using T-duality, and found correspondences between string field theory
solutions and effective field theory solutions. Our analysis is in the
closed bosonic string field theory and the heterotic string field theory.
However, since our analysis has used little information on interaction terms,
our result seems universal. Although consistent type II closed string field
theory has not been constructed yet, our analysis can be applied to it once
it is constructed and will yield essentially the same result.

Terms in the field redefinitions left undetermined in section 4 probably
depend on the definition of interaction terms and the gauge fixing condition
for massive modes, and can be determined by
direct application of the integrating-out procedure. Higher derivative
terms can also be determined by computing $\ap$-correction for the
T-duality rule in the effective theory. It is important to determine
at least nonderivative terms for understanding how to extract physical
information from string field theory.

In this paper we have not paid much attention to the tachyon component.
It is interesting to investigate field redefinition for the tachyon
to understand what happens when closed string tachyon is condensed.
For example, in a study of tachyon condensation in orbifolds\cite{oz2},
it has been found that both twisted tachyon and untwisted massless
components in the string field are involved in the tachyon potential. 
We can see if it is also true in terms of the redefined variables
and if the conjecture in \cite{dbhlkr} is strictly true.
We also have not taken into account the gauge field in the heterotic
theory. We can construct solutions with nonzero gauge field, which
have properties similar to the solutions in this paper by the same
mechanism. Our discussion on T-duality is also extended to the case
with the gauge field.

The R-sector part of the heterotic string field theory has not been
constructed yet, and therefore we do not know how supersymmetry is
incorporated in this theory. Since some of our solutions are expected to be
supersymmetric, it is desirable to construct supersymmetry transformation and 
confirm that our solutions leave some supersymmetry unbroken.

%%%%%%%%%%%%%%%%%%%%%%%%%%%%%%%%%%%%%%%%%%%%%%%%%%%%%%%%%%%%%%%
\vs{.5cm}
\noindent
{\large\bf Acknowledgments}\\[.2cm]
The author wishes to thank M.\ Headrick and Y.\ Okawa for useful discussion
and B.\ Zwiebach for helpful comments on the manuscript.
This work is supported in part by funds provided by the U.S. Department of
Energy (D.O.E.) under cooperative research agreement DF-FC02-94ER40818,
and by the Nishina Memorial Foundation.
%%%%%%%%%%%%%%%%%%%%%%%%%%%%%%%%%%%%%%%%%%%%%%%%%%%%%%%%%%%%%%%

%%%%%%%%%%%% References %%%%%%%%%%%%%%%%%%%%%%%%%
\newcommand{\J}[4]{{\sl #1} {\bf #2} (#3) #4}
\newcommand{\andJ}[3]{{\bf #1} (#2) #3}
\newcommand{\AP}{Ann.\ Phys.\ (N.Y.)}
\newcommand{\MPL}{Mod.\ Phys.\ Lett.}
\newcommand{\NP}{Nucl.\ Phys.}
\newcommand{\PL}{Phys.\ Lett.}
\newcommand{\PR}{Phys.\ Rev.}
\newcommand{\PRL}{Phys.\ Rev.\ Lett.}
\newcommand{\PTP}{Prog.\ Theor.\ Phys.}
\newcommand{\hepth}[1]{{\tt hep-th/#1}}
%%%%%%%%%%%%%%%%%%%%%%%%%%%%%%%%%%%%%%%%%%%%%%%%

\end{document}